\documentclass[10pt]{article}

\usepackage{amsmath}
\usepackage{amssymb}

\usepackage{graphicx}

\usepackage{cite}
\usepackage{color}

\usepackage										{graphicx}
\graphicspath										{./figures/}
\usepackage										{amssymb}
\usepackage										{amsmath}
\usepackage										{url}
\usepackage		[numbers, 	sort]					{natbib}
\usepackage		[font=small,labelfont=bf]				{caption}
\usepackage										{indentfirst}
\usepackage										{microtype}
\usepackage										{cleveref}
\usepackage		[]								{booktabs}
\crefname{equation}{Eq.}{Eqs.}
\crefname{figure}{Fig.}{Figs.}
\Crefname{figure}{Figure}{Figures}
\providecommand{\abs}[1]{\left\lvert#1\right\rvert}
\providecommand{\mean}[1]{\left\langle#1\right\rangle}
\providecommand{\norm}[1]{\left\lVert#1\right\rVert}

\usepackage[labelfont=bf,labelsep=period,justification=raggedright]{caption}

\makeatletter
\renewcommand{\@biblabel}[1]{\quad#1.}
\makeatother

\date{}

\begin{document}

\begin{flushleft}
{\Large
\textbf{Resolving Structure in Human Brain Organization: Identifying Mesoscale Organization in Weighted Network Representations}
}
\\
Christian Lohse$^{1}$,
Danielle S. Bassett$^{2,3,4,\ast}$,
Kelvin Lim$^{5}$
Jean M. Carlson$^{2}$
\\
\bf{1} Kirchhoff Institute for Physics, University of Heidelberg, Germany 69120
\\
\bf{2} Department of Physics, University of California, Santa Barbara, CA 93106, USA
\\
\bf{3} Sage Center for the Study of the Mind, University of California, Santa Barbara, CA 93106, USA
\\
\bf{4} Department of Bioengineering, University of Pennsylvania, Philadelphia, PA 19104, USA
\\
\bf{5} Department of Psychiatry, University of Minnesota, Minneapolis, MN 55455, USA
\\
$\ast$ E-mail: Corresponding dsb@seas.upenn.edu
\end{flushleft}

\newpage

\section*{Abstract}
Human brain anatomy and function display a combination of modular and hierarchical organization, suggesting the importance of both cohesive structures and variable resolutions in the facilitation of healthy cognitive processes. However, tools to simultaneously probe these features of brain architecture require further development. We propose and apply a set of methods to extract cohesive structures in network representations of brain connectivity using multi-resolution techniques. We employ a combination of soft thresholding, windowed thresholding, and resolution in community detection, that enable us to identify and isolate structures associated with different weights. One such \emph{mesoscale} structure is bipartivity, which quantifies the extent to which the brain is divided into two partitions with high connectivity between partitions and low connectivity within partitions. A second, complementary mesoscale structure is modularity, which quantifies the extent to which the brain is divided into multiple communities with strong connectivity within each community and weak connectivity between communities. Our methods lead to multi-resolution curves of these network diagnostics over a range of spatial, geometric, and structural scales. For statistical comparison, we contrast our results with those obtained for several benchmark null models. Our work demonstrates that multi-resolution diagnostic curves capture complex organizational profiles in weighted graphs. We apply these methods to the identification of resolution-specific characteristics of healthy weighted graph architecture and altered connectivity profiles in psychiatric disease.


%

%

\section{Introduction}\label{sec:introduction}

Noninvasive neuroimaging techniques provide quantitative measurements of structural and functional connectivity in the human brain. Functional magnetic resonance imaging (fMRI) indirectly resolves time dependent neural activity by measuring the blood-oxygen-level-dependent (BOLD) signal while the subject is at rest or performing a cognitive task. Diffusion weighted imaging (DWI) techniques use MRI to map the diffusion of water molecules along white matter tracts in the brain, from which anatomical connections between brain regions can be inferred. In each case, measurements can be represented as a weighted network \cite{Bassett2006a,Bassett2006b,Bullmore2009,Bullmore2011,Kaiser2011,Sporns2011,Stam2012}, in which nodes correspond to brain regions, and the weighted connection strength between two nodes can, for example, represent correlated activity (fMRI) or fiber density (DWI). The resulting network is complex, and richly structured, with weights that exhibit a broad range of values, reflecting a continuous spectrum from weak to strong connections.

The network representation of human brain connectivity facilitates quantitative and statistically stringent investigations of human cognitive function, aging and development, and injury or disease. Target applications of these measurements include disease diagnosis, monitoring of disease progression, and prediction of treatment outcomes \cite{Bassett2009,Fornito2012a,Fornito2012b,He2009,He2010}. However, efforts to develop robust methods to reduce these large and complex neuroimaging data sets to statistical diagnostics that differentiate between patient populations have been stymied by the dearth of methods to quantify the statistical significance of apparent group differences in network organization\cite{Wijk2010,Ginestet2011,Bassett2011a,Fornito2013}.

Network comparisons can be performed in several ways. In one approach, network comparisons are made after applying a threshold to weighted structural and functional connectivity matrices to fix the number of edges at a constant value in all individuals \cite{Bullmore2011,Wijk2010}. Edges with weights passing the threshold are set to a value of $1$ while all others are set to a value of $0$ (a process referred to as `binarizing'). In some cases results are tested for robustness across multiple thresholds, although this increases the probability of Type I (false positive) errors from multiple non-independent comparisons. More generally, this procedure disregards potentially important neurobiological information present in the original edge weights. A second approach involves examination of network geometry in the original weighted matrices without binarizing. However, because the values of weighted metrics can be influenced by both the average weight of the matrix and the distribution of weights, this approach presents peculiar complications for the assessment of group differences \cite{Ginestet2011,Bassett2011a}. Critically, neither of these two approaches for network comparison allow for a principled examination of network structure as a function of weight (strong versus weak connections) or space (short versus long connections). Disease-related group differences in network architecture that are present only at a particular edge weight range or at a specific spatial resolution can therefore remain hidden.

\begin{table}[!b]
\centering
	\includegraphics[width = \textwidth]{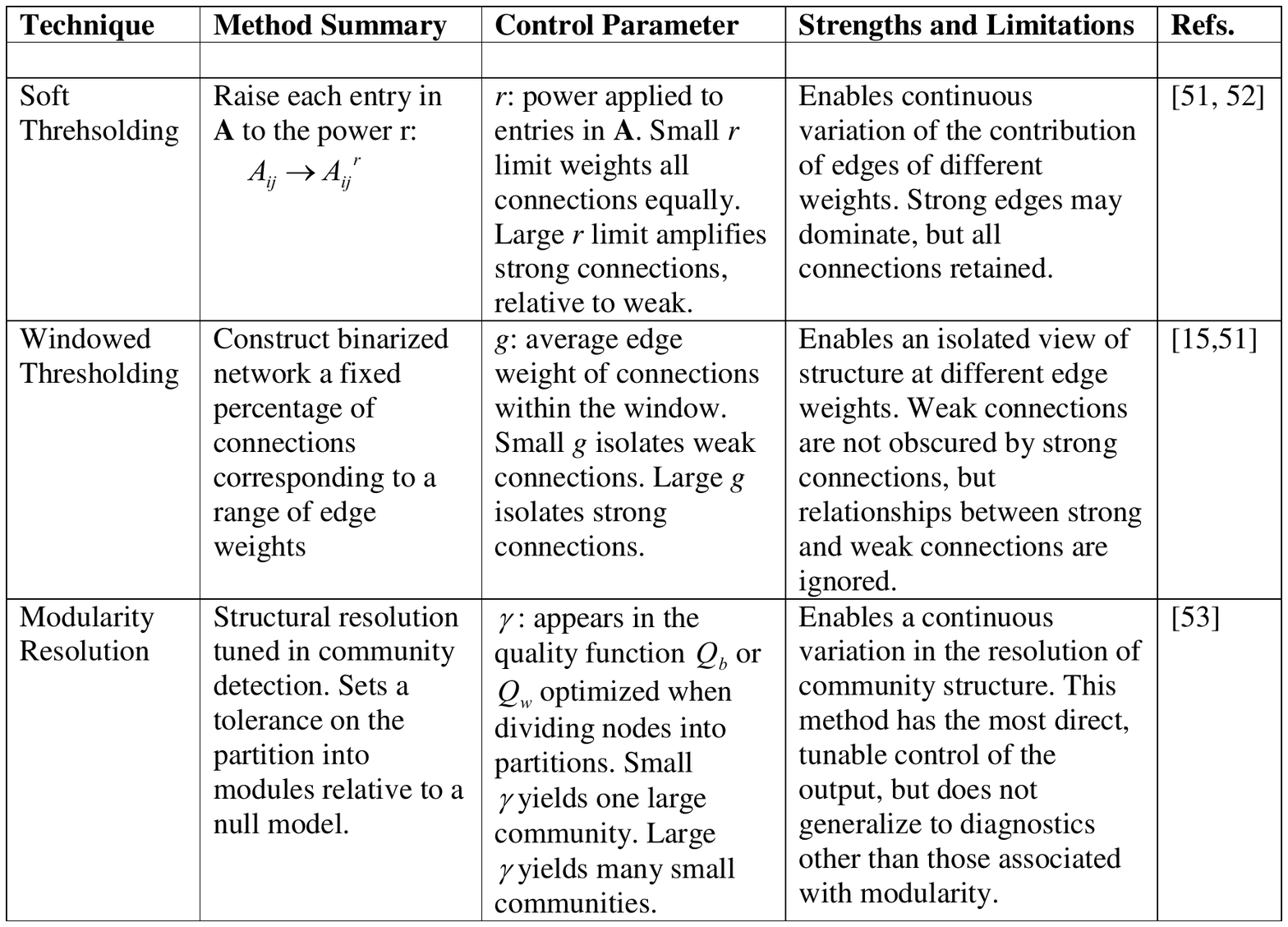}
	\caption{Summary of multi-resolution methods for network diagnostics. We use soft thresholding, windowed thresholding, and variation in the resolution parameter of modularity maximization to probe network architecture across scales (\emph{Column 1}). For each approach, we provide a method summary (\emph{Column 2}), a description of the control parameter (\emph{Column 3}), a brief synopsis of the strengths and limitations of the approach (\emph{Column 4}), and a few relevant references (\emph{Column 5}).}
\label{table:methods_summary}
\end{table}

In this paper, we employ several techniques to examine the multi-resolution structure of weighted connectivity matrices: soft thresholding, windowed thresholding, and modularity resolution. A summary of these techniques is presented in Table I, and each method is discussed in more detail in the Methods section. We apply these techniques to structural networks extracted from diffusion tractography data in five healthy human subjects (N=5) \cite{Hagmann2008} and functional networks extracted from resting state fMRI data in people with schizophrenia and healthy controls (N=58) \cite{Bassett2012a}. As benchmark comparisons, we also explore a set of synthetic networks that includes a random Erd\H{o}s-R\'{e}nyi network (ER), a ring lattice (RL), a small-world network (SW), and a fractal hierarchical network (FH).

While multi-resolution techniques could be usefully applied to a broad range of network diagnostics, here we focus on two complementary mesoscale characteristics that can be used to probe the manner in which \emph{groups} of brain regions are connected with one another: modularity and bipartivity. Modularity quantifies community structure in a network by identifying groups of brain regions that are more strongly connected to other regions in their group than to regions in other groups \cite{Porter2009,Fortunato2010}. Communities of different sizes, nested within one another, have been identified in both structural and functional brain networks \cite{Bassett2010c,Gallos2012,Meunier2009,Meunier2010,Zhou2006} and are thought to constrain information processing \cite{Wang2011,Kaiser2010,Rubinov2011b}. Bipartivity quantifies bipartite structure in a network by separating brain regions into two groups with sparse connectivity within each group and dense connectivity between the two groups. The dichotomous nature of bipartitivity is particularly interesting to quantify in systems with bilateral symmetry such as the human brain, in which inter- and intra-hemispheric connectivity display differential structural \cite{Klimm2013} and functional \cite{Doron2012} network properties.

The rest of this paper is organized as follows. In Methods (Section 2) we describe the empirical brain networks as well as the synthetic networks used in this study. This is followed by a description of the network properties measured, and the multi-resolution techniques that identify how these properties vary across different connection properties in the networks. In Results (Section 3) we apply these methods to the empirical and synthetic networks to characterize multi-resolution modular and bipartite structure. We illustrate the utility of these approaches in identifying characteristics such as community laterality and radius that are peculiar to the patterns of weak versus strong connections and short versus long connections, and the modulation of these functions by disease. We conclude in Section 4 with a Discussion of our results. Additional methodological considerations are presented as Supplementary Information.

\section{Methods}
\label{sec:methods}
This section has three major components: (1) a description of the empirical data examined in this study as well as the simpler synthetic networks employed to illustrate our techniques in well-characterized, controlled settings and to provide benchmark null tests for comparison with brain data, (2) a summary of the graph diagnostics used in our analysis to examine properties of mesoscale network architecture, and (3) definitions of the soft thresholding, windowed thresholding, and modularity resolution techniques which provide a means to resolve network structure at different connection strengths.

\subsection{Empirical and Synthetic Benchmark Networks}
We examine two separate neuroimaging data sets acquired noninvasively from humans. The first is a set of $N=6$ structural networks constructed from diffusion spectrum imaging (DSI) data and the second is a set of $N=58$ functional networks constructed from resting state functional magnetic resonance imaging (fMRI) data. For comparison, we generate 4 types of synthetic network models that range from an Erd\H{o}s-R\'{e}nyi random graph to models that include more complex structure, including hierarchy and clustering.

Each network is described by an adjacency matrix $\mathbf{A}$ whose $ij^{\mathrm{th}}$ entry describes the weight of the edge connecting nodes $i$ and $j$. For the empirical brain networks, the edge weights are determined by neuroimaging measurements. For the synthetic networks, we construct weights to mimic the structural organization of the network, as described below.

\subsubsection{Empirical Brain Networks: Data Acquisition and Preprocessing }
\label{sec:empirical.networks}
\label{sec:methods.data}

\paragraph{Structural Networks:} We construct an adjacency matrix $\mathbf{A}$ whose $ij^{\mathrm{th}}$ entries are the probabilities of fiber tracts linking region $i$ with region $j$ as determined by an altered path integration method. The resultant adjacency matrix $\mathbf{A}$ contains entries $A_{ij}$ that represent connection weights between the 998 regions of interest extracted from the whole brain. We also define the distance matrix $\mathbf{L}$ to contain entries $L_{ij}$ that are the average curvilinear distance traveled by fiber tracts that link region $i$ with region $j$, or the Euclidean distance between region $i$ and $j$, where no data of the arc length was available. We define separate $\mathbf{A}$ and $\mathbf{L}$ adjacency matrices for each of 5 healthy adults with 1 adult scanned twice (treated as 6 separate scans throughout the paper). For further details, see \cite{Hagmann2008}.

\paragraph{Functional Networks:} We construct an adjacency matrix $\mathbf{A}$ whose $ij^{\mathrm{th}}$ entry is given by the absolute value of the Pearson correlation coefficient between the resting state wavelet scale 2 (0.06--0.12 Hz) blood oxygen level dependent (BOLD) time series from region $i$ and from region $j$. The resultant adjacency matrix $\mathbf{A}$ contains entries $A_{ij}$ that represent functional connection weights between 90 cortical and subcortical regions of interest extracted from the whole brain automated anatomical labeling (AAL) atlas \cite{Tzourio2002}.

We define separate adjacency matrices for each of 29 participants with chronic schizophrenia (11 females; age 41.3 $\pm$ 9.3 (SD)) and 29 healthy participants (11 females; age 41.1 $\pm$ 10.6 (SD)) (see \cite{Camchong2009} for detailed characteristics of participants and imaging data). We note that the two groups had similar mean RMS motion parameters (two-sample t-tests of mean RMS translational and angular movement were not significant at $p=0.14$ and $p=0.12$, respectively), suggesting that identified group difference in network properties could not be attributed to differences in movement during scanning.

\subsubsection{Synthetic Benchmark Networks}
\label{sec:methods.models}

\begin{figure}[!ht]
\centering
	\includegraphics[width = \textwidth]{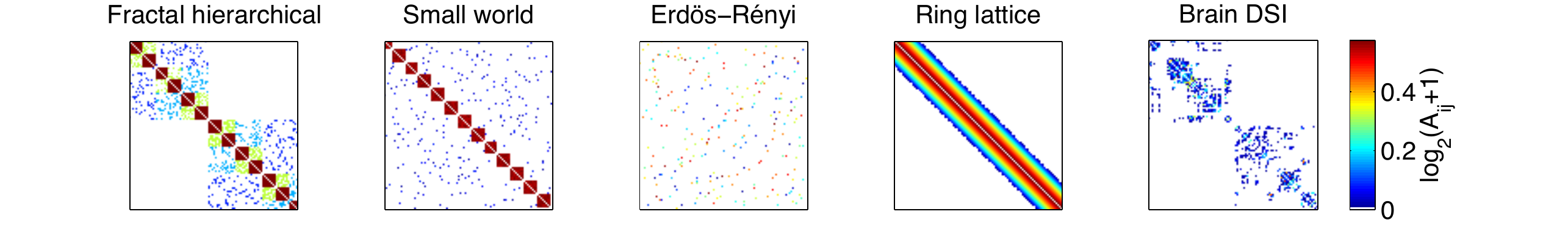}
	\caption{Weighted connection matrices of the data sets. The topologies change as a function of weight, particularly for the fractal hierarchical and the small world network. Here we illustrate the connections between the nodes 450 to 550 (of 1000 total nodes). The windowed thresholding technique isolates topological characteristics in the subnetworks of nodes of similar weight (color) in the adjacency matrix. We report the initial benchmark results for a window size of 25\% but find that results from other window sizes are qualitatively similar (see the Supplementary Information).
	}
	\label{fig:adj_mat}
\end{figure}

In addition to empirical networks, we also examine 4 synthetic model networks: an Erd\"os-R\'enyi random network, a ring-lattice, a small world network, and a fractal hierarchical network (see Fig.~\ref{fig:adj_mat}). All networks were created using modified code from \cite{Rubinov2010}.

The Erd\"os-R\'enyi and the ring-lattice networks are important benchmark models used in a range of contexts across a variety of network systems. Small world and fractal hierarchical networks incorporate clustering and layered structure, reminiscent of properties associated with brain networks \citep{Sporns2006}. We emphasize that the synthetic models are not intended to be realistic models of the brain. Rather, we use them to isolate structural drivers of network topology and to illustrate the utility of multi-resolution approaches \cite{Onnela2012} in a controlled setting.

Most synthetic network models, including all those that we study in this paper were originally developed as binary graphs (i.e. all edge weights equal to either $0$ or $1$). Since we are specifically interested in the effect of edge weights on network properties, we consider weighted generalizations of these models, in which the weights of edges are chosen to maintain essential structural properties of the underlying graph (see Figure~\ref{fig:adj_mat}A).

\paragraph{Erd\"os-R\'enyi Random (ER) Model:}
The Erd\"os-R\'enyi random graph serves as an important benchmark null model against which to compare other empirical and synthetic networks. The graph $G(N,K)$ is constructed by assigning a fixed number of connections $K$ between the $N$ nodes of the network. The edges are chosen uniformly at random from all possible edges, and we assign them weights drawn from a uniform distribution in $[0,1]$.

\paragraph{Ring-Lattice (RL) Model:} The one-dimensional ring-lattice model can be constructed for a given number of edges $K$ and number of nodes $N$ by first placing edges between nodes separated by a single edge, then between pairs of nodes separated by $2$ edges and so on, until all $K$ connections have been assigned. To construct a geometric (i.e. weighted) chain-like structure, we assign the weights of each edge to mimic the topological structure of the binary chain lattice. Edge weights decreased linearly from $1$ to $0$ with increasing topological distance between node pairs.

\paragraph{Small World (SW) Model:}
\label{sec:methods.models.sw}
Networks with high clustering and short path lengths are often referred to as displaying ``small world'' characteristics \citep{Milgram1969,Watts1998}. Here we construct a modular small world model \cite{Sporns2006} by connecting $N=2^n$ nodes in elementary groups of fully connected clusters of $M=2^m$ nodes, where $m$ and $n$ are integers. If the number of edges in this model is less than $K$, the remaining edges are placed uniformly at random throughout the network. The edges in this model are either intra-modular (placed within elementary groups) or inter-modular (placed between elementary groups). To construct a geometrical modular small world model, we assigned different weights to the two types of edges: $A_{ij} = 1$ for intra-modular edges and $A_{ij} = 0.5$ for inter-modular edges.

To ensure our comparisons are not dominated by the overall network density, we construct synthetic network models with $N$ nodes and $K$ edges to match the empirical data \citep{Wijk2010}. For modular small world networks, this stipulation requires that we modify the algorithm above to produce networks in which the number of nodes is not necessarily an exact power of two. For comparisons to the structural (functional) brain data, we generate a model as described above with $N=1024, M=8$ ($N = 128, M=4$) and then chose $24$ ($38$) nodes at random, which are deleted along with their corresponding edges.

\paragraph{Fractal Hierarchical (FH) Model:}
\label{sec:methods.models.fr}
Network structure in the brain displays a hierarchically modular organization \cite{Bassett2010c,Gallos2012,Meunier2009,Meunier2010,Zhou2006} thought generally to constrain information processing phenomena \cite{Wang2011,Kaiser2010,Rubinov2011b}. Following \cite{Sporns2006,Ravasz2003}, we create a simplistic model consisting of $N=2^n$ nodes that form $l_{max}\leq n$ hierarchical levels to capture some essential features of fractal hierarchical geometry in an idealized setting. At the lowest level of the hierarchy, we form fully connected elementary groups of size $M = 2^m, m=n-l_{max}+1$ and weight all edges with the value $1$. At each additional level of the hierarchy $l$, we place edges uniformly at random between two modules from the previous level $l-1$. The density of these inter-modular edges is given by the probability $p (l)=E^{-l}$ where $E$ is a control parameter chosen to match the connection density $\frac{K}{N(N-1)}$ of the empirical data set. To mimic this defining topological feature, we construct the network geometry by letting edge weights at level $l$ equal $p(l)$.
For comparisons to the structural (functional) brain data, we generate a model as described above with $N=1024, M=8$ $l_{max}=7$ ($N=128, M=4, l_{max}=7$) and then chose $24$  ($38$) nodes at random, which are deleted along with their corresponding edges.

\paragraph{Constructing Network Ensembles:} Each empirical data set displays variable $K$ values. To construct comparable ensembles of synthetic networks, we drew $K$ from a Gaussian distribution defined by the mean and standard deviation of $K$ in the empirical data. For comparisons to DSI empirical data, we constructed ensembles of $N = 1000, K = (2.88 \pm 0.10) \cdot 10^4$ composed of 50 realizations per model. See Supplementary Information Table S1 for details on the ensembles.

\subsection{Mesoscale Network Diagnostics}
In this paper we focus on two, complementary network characteristics that can be used to probe the manner in which \emph{groups} of brain regions are connected with one another: modularity and bipartivity. These methods, however, are more generally applicable, and could be used to evaluate weight dependence of other metrics and data sets as well\cite{Newman2010}.

\subsubsection{Community Detection and Modularity}
\label{sec:methods.modularity}
Community detection by modularity maximization identifies a partition of the nodes of a network into groups (or \emph{communities}) such that nodes are more densely connected with other nodes in their group than expected in an appropriate statistical null model. In this paper, we use this method to extract values of network modularity and other related diagnostics including community laterality, community radius, and the total number and sizes of communities. An additional feature of modularity maximization that is particularly useful for our purposes is its resolution dependency, which enables us to continuously monitor diagnostic values over different organizational levels in the data.

Community detection can be applied to both binary and weighted networks.
Assuming node $i$ is assigned to community $c_{i}$ and node $j$ is assigned to community $c_{j}$, the quality of a partition for a binary network is defined using a modularity index \cite{markfast,Newman2004,Newman2006b,Porter2009,Fortunato2010}:
\begin{equation}
Q_{b} = \frac{1}{2K}\sum_{i,j}(B_{ij}-\gamma P_{ij})\delta (c_i,c_j),
\label{eq:def_mod_b}
\end{equation}
where $\frac{1}{2K}$ is a normalization constant, $\mathbf{B}$ is the binary adjacency matrix, $\gamma$ is a structural resolution parameter, and $P_{ij}$ is the probability of a connection between nodes $i$ and $j$ under a given null model. Here we use the Newman-Girvan null model \cite{Newman2004} that defines the connection probability between any two nodes under the assumption of randomly distributed edges and the constraint of a fixed degree distribution: $P_{ij} = \frac{k_i k_j}{2K}$, where $k_i$ is the degree of node $i$. We optimize $Q_{b}$ from Equation 1 using a Louvain-like \cite{Blondel2008} locally greedy algorithm \cite{genlouvain} to identify the optimal partition of the network into communities.

For weighted networks, a generalization of the modularity index is defined as \cite{Traag2009,Arenas2008}:
\begin{equation}
Q_{w} = \frac{1}{2W}\sum_{i,j}(A_{ij}-\gamma \frac{s_i s_j}{2W})\delta (c_i,c_j),
\label{eq:def_mod_w}
\end{equation}
where $W=\frac{1}{2}\sum_{ij}A_{ij}$ is the total edge weight, $\mathbf{A}$ is the weighted adjacency matrix, and $s_i$ is the strength of node $i$ defined as the sum of the weights of all edges emanating from node $i$.

Due to the non-unique, but nearly-degenerate algorithmic solutions for $Q_{b}$ and $Q_{w}$ obtained computationally \cite{Good2010}, we perform 20 optimizations of Equation 1 or 2 for each network under study. In the results, we report mean values of the following $4$ diagnostics over these optimizations: the modularity $Q_{b}$ or $Q_{w}$, the number of  communities, the number of singletons (communities that consist of only a single node), and the community laterality and radius (defined in a later sections). We observed that the optimization error in these networks is significantly smaller than the empirical inter-individual or synthetic inter-realization variability, suggesting that these diagnostics produce reliable measurements of network structure (see Supplementary Information).

\paragraph{Community Laterality:}
\label{sec:methods.laterality}
Laterality is a property that can be applied to any network in which each node can be assigned to one of two categories, and within the community detection method, describes the extent to which a community localizes to one category or the other \cite{Doron2012}. In neuroscience, an intuitive category is that of the left and right hemispheres, and in this case laterality quantifies the extent to which the identified communities in functional or structural brain networks are localized within a hemisphere or form bridges between hemispheres.

For an individual community $c$ within the network, the laterality $\Lambda_c$ is defined as \citep{Doron2012}:
\begin{equation}
\Lambda_c = \frac{\abs{N_r - N_l}}{N_c},
\end{equation}
where $N_r$ and $N_l$ are the number of nodes located in the left and right hemispheres respectively, or more generally in one or other of the two categories, and $N_c$ is the total number of nodes in $c$. The value of $l$ ranges between zero (i.e., the number of nodes in the community are evenly distributed between the two categories) and unity (i.e., all nodes in the community are located in a single category).

We define the laterality of a given partition of a network as:
\begin{equation}
\Lambda = \frac{1}{N}\left(\sum_c N_c \Lambda_c - \mean{\sum_c N_c \Lambda_c}\right),
\label{eq:def_lat}
\end{equation}
where we use $\mean{\sum_c N_c \Lambda_c}$ to denote the expectation value of the laterality under the null model specified by randomly reassigning nodes to the two categories while keeping the total number of nodes in each category fixed. We estimate the expectation value by calculating $\sum_c N_c \Lambda_c$ for 1000 randomizations of the data, which ensures that the error in the estimation of the expectation value is of order $10^{-3}$.

The laterality of each community is weighted by the number of nodes it contains, and the expectation value is subtracted to minimize the dependence of $\Lambda$ on the number and sizes of detected communities. This correction is important for networks like the brain that exhibit highly fragmented structure. Otherwise the estimation would be biased by the large number of singletons, that by definition have a laterality of $\Lambda_c=1.0$.

\paragraph{Community Radius:}
\label{sec.methods.rho}
To measure the spatial extent of a community in a physically embedded network such as the human brain, we define the community `radius' $\rho_c$ as the standard deviation of the lengths of the position vectors of all nodes in the community:
\begin{equation}
	\rho_c = \frac{1}{N_c}( \sum_{i \in c} \norm{\mathbf{r}_i}^2 - \lVert\sum_{i \in c} \mathbf{r}_i \rVert ^2)^\frac{1}{2} ,
\label{eq:def_rho_c}
\end{equation}
where $N_c$ is the number of nodes in community $c$ and $\mathbf{r}_i$ is the position vector of node $i$.

The average community radius of the entire network $\mathbf{A}$ is
\begin{equation}
	\rho_A = \frac{1}{N} \sum_c N_c \frac{\rho_c}{R_A} ,
\label{eq:def_rho}
\end{equation}
where $N_{c}$ serves to weight every community by the number of nodes it contains (compare to Equation 4), and $R_{A}$ is a normalization constant equal to the `radius' of the entire network: $R_A = \frac{1}{N} \sqrt{(\sum_{i=1}^{N} \norm{\mathbf{r}_i}^2 - \lVert\sum_{i = 1}^{N} \mathbf{r}_i \rVert ^2)}$. In this investigation, community radius is evaluated only for the empirical brain networks, since the synthetic networks lack a physical embedding.

\subsubsection{Bipartivity}
\label{sec:methods.bipartivity}
Bipartivity is a topological network characteristic that occurs naturally in certain complex networks with two types of nodes. A network is perfectly bipartite if it is possible to separate the nodes of the network into two groups, such that all edges lie between the two groups and no edges lie within a group. In this sense bipartivity is a complementary diagnostic to modularity, which maximizes the number of edges within a group and minimizes the number of edges between groups.

Most networks are not perfectly bipartite, but still show a certain degree of bipartivity. A quantitative measure of bipartivity can be defined by considering the subgraph centrality $\mean{SC}$ of the network, which is defined as a weighted sum over the number of closed walks of a given length. Because a network is bipartite if and only if there only exist closed walks of even length, the ratio of the contribution of closed walks of even length to the contribution of closed walks of any length provides a measure of the network bipartivity \citep{Estrada2005}. As shown in \citep{Estrada2005}, this ratio can be calculated from the eigenvalues $\lambda_{j}$ of the adjacency matrix \cite{Estrada2005}:
\begin{equation}\label{eq:def_beta}
\beta =  \frac{\mean{SC_{even}}}{\mean{SC}} = \frac{\sum_{\lambda_j} \cosh(\lambda_j)}{\sum_{\lambda_j}\exp(\lambda_j)} ,
\end{equation}
where $\mean{SC_{even}}$ is the contribution of closed walks of even length to the subgraph centrality. In this formulation, $\beta$ takes values between $0.5$ for the least bipartite graphs (fully connected graphs) and $1.0$ for a perfectly bipartite graph \citep{Estrada2005}.

The definition of bipartivity given in Equation 7 provides an estimate of $\beta$ but does not provide a partition of network nodes into the two groups. To identify these partitions, we calculate the eigenvector corresponding to the smallest eigenvalue $\lambda_\mathrm{min}$ of the modularity quality matrix $M_{ij} = \frac{1}{2K}(A_{ij}- \frac{k_i k_j}{2K})$ \cite{Newman2006}. We then partition the network according to the signs of the entries of this eigenvector: nodes with positive entries in the eigenvector form group one and nodes with negative entries in the eigenvector form group two. This spectral formulation demonstrates that bipartivity is in some sense an anti-community structure \cite{Newman2006}: community structure corresponds to the highest eigenvalues of $\mathbf{M}$ while the bipartivity corresponds to the lowest eigenvalue of $\mathbf{M}$.

\subsection{Multi-resolution Methods}

We quantify organization of weighted networks across varying ranges of connection weights (weak to strong) and connection lengths (short to long), using three complementary approaches: soft thresholding, windowed thresholding, and multi-resolution community detection. Each method employs a control parameter that we vary to generate network diagnostic curves, representing characteristics of the network under study. We refer to these curves as \emph{mesoscopic response functions} (MRF) of the network \cite{Onnela2010}. A summary of the methods and control parameters is contained in Table I.

\subsubsection{Soft Thresholding}
\label{sec:methods.power_and_rew}
To date most investigations have focused on cumulative thresholding \cite{Bullmore2011}, where a weighted network is converted to a binary network (binarized) by selecting a single threshold parameter $\tau$ and setting all connections with weight $A_{ij}>\tau$ to $1$ and all other entries to $0$. Varying $\tau$ produces multiple binary matrices over a wide range of connection densities (sparsity range) on which the network diagnostic of interest can be computed. One disadvantage of this \emph{hard} thresholding technique is that it is predicated on the assumption that matrix elements with weights just below $\tau$ are significantly different from weights just above $\tau$.

Soft thresholding \cite{Schwarz2011,Zhang2005} instead probes the full network geometry as a function of edge weight. We create a family of graphs from each network by reshaping its weight distribution\footnote{Note that we normalize edge weights prior to this procedure to ensure that all $A_{ij}$ are in $[0,1]$.}: $A_{ij} \to A_{ij}^r$. As we vary $r$ from $0$ to $\infty$, the weight distribution shifts accordingly. At $r=0$, the weight distribution approximates a delta function: all nonzero elements of the resulting matrix have the same weight of $A_{ij}^{r}=1$, which corresponds to the application of a hard threshold at $\tau=0$. At $r\to \infty$, the weight distribution is heavy-tailed: the majority of elements of the resulting matrix are equal to $0$, corresponding to the application of a hard threshold at $\tau \to \max_{ij} A_{ij}$.

\subsubsection{Windowed Thresholding}
\label{sec:methods.thresholding}
A potential disadvantage of both the cumulative and soft thresholding approaches is that results may be driven by effects of both connection density and the organization of the edges with the largest weights. As a result, these procedures can neglect underlying structure associated with weaker connections.

Windowed thresholding \cite{Bassett2011a, Schwarz2011} instead probes the topology of families of edges of different weights independently. We replace the cumulative threshold $\tau$ with a threshold window $[g_{-},g_{+}]$ that enforces an upper and lower bound on the edge weights retained in the construction of a binary matrix. The window size is specified as a fixed percentage $p$ of connections which are retained in each window, and the average weight $g$ of connections in each window is the variable control parameter. Here $g, g_{-}$ and $g_{+}$ are given as percentiles of all connections.

For each network $\mathbf{A}$, we construct a family of binary graphs $\mathbf{B}^p(g)$, each of which depends on the window size $p$ (percentage of connections retained), and the average weight $g$ of connections in the window (together $p$ and $g$ determine the upper and lower bounds, $g_{-}$ and $g_{+}$, defining the window):
\begin{equation}
B^{p}_{ij} (g) =
\begin{cases}
	1, & \mbox{if $A_{ij} \in [g_{-}, g_{+}]$;} \\
	0, & \mbox{otherwise.}
\end{cases}
\label{eq:binarize}
\end{equation}
The fixed window size (5\%-25\%) mitigate biases associated with variable connection densities \cite{Bassett2011a,Bullmore2011,Ginestet2011}. Each resulting binary graph $\mathbf{B}^p(g)$ in the family summarizes the topology of edges with a given range of weights. The window size sets a resolution limit to the scale on which one can observe weight dependent changes in structure (see the Supplementary Information).

To probe the roles of weak versus strong edges and short versus long edges in the empirical brain networks, we extract families of graphs based on (i) networks weighted by correlation strength (fMRI), (ii) networks weighted by number of white matter streamlines (DSI), (iii) networks weighted by Euclidean distance between brain regions (DSI), and (iv) one network weighted by fiber tract arc length (DSI). For each case we compute graph diagnostics as a function of the associated connection weight $g$. This method isolates effects associated with different connection weights, but does not resolve network organization across different ranges of weights.

\subsubsection{Resolution in Community Detection}
\label{sec:methods.gamma}
The definition of modularity in Equations 1,2 includes a resolution parameter $\gamma$ \citep{Fortunato2006,Reichardt2006,Reichardt2007} which tunes the size of communities in the optimal partition: high values of $\gamma$ lead to many small communities and low values of $\gamma$ lead to a few large communities. We vary $\gamma$ to explore partitions with a range of mean community sizes: for the smallest value of $\gamma$ the partition contains a single community of size $N$, while for the largest value of $\gamma$ the partition contains $N$ communities of size unity. In physically embedded systems, such as brain networks, one can probe the relationship between structural scales uncovered by different $\gamma$ values and spatial scales defined by the mean community radius $\rho$.

\section{Results}
\label{sec:results}

\begin{figure}[!b]
\centering
	\includegraphics[width = \textwidth]{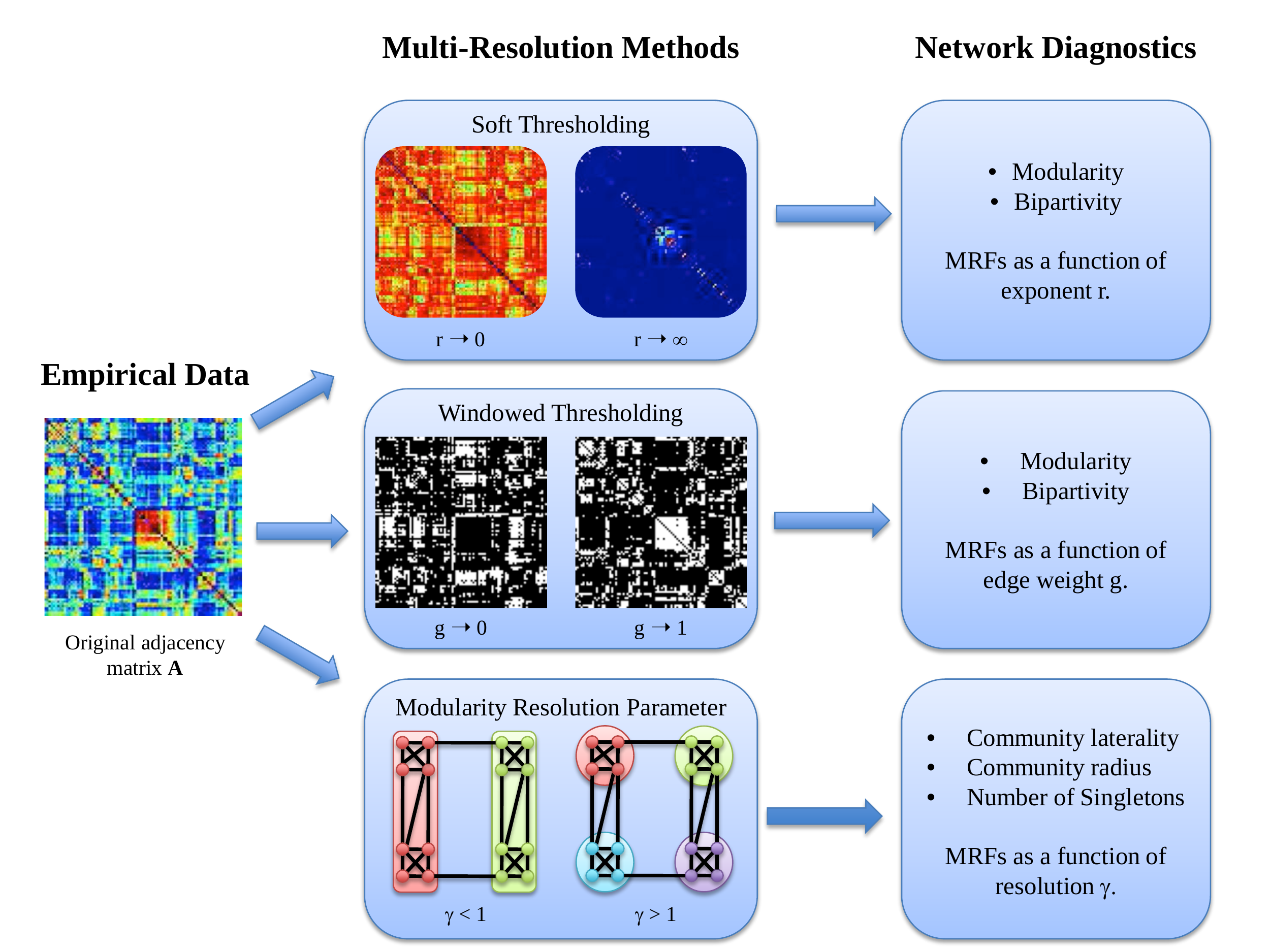}
	\caption{Pictorial Schematic of Multi-Resolution Methods for Weighted Networks. We can apply soft or windowed thresholding, and vary the resolution parameter of modularity maximization to uncover multiresolution structure in empirical data that we summarize in the form of MRFs of network diagnostics.}
\label{fig:flowchart}
\end{figure}

In this section we apply multi-resolution diagnostics to a sequence of $5$ problems involving weighted empirical and synthetic network data (see Fig.~\ref{fig:flowchart}). We summarize these problems below.
\begin{itemize}
\item \textbf{Problem 1: Probing Drivers of Weighted Modularity} We begin our analysis with the weighted modularity $Q_w$ (Equation 2), a diagnostic that is based on the full weighted adjacency matrix. Using soft thresholding, we show that $Q_w$ is most sensitive to the strongest connections. This motivates the use of windowed thresholding to isolate the topology of connections based on their weight.
\item \textbf{Problem 2: Determining Network Differences in Multiresolution Structure} In anatomical brain data and corresponding synthetic networks, we use windowed thresholding to obtain multi-resolution response functions (MRFs) for modularity $Q_b(g)$ and bipartivity $\beta (g)$ as a function of the average weight $g$ within the window. MRFs of synthetic networks do not resemble MRFs of the brain.
\item \textbf{Problem 3: Uncovering Differential Structure in Edge Strength \& Length} Using windowed thresholding, we probed the multiresolution structure in anatomical brain data captured by two measures of connection weight: fiber density and fiber length. We observe that communities involving short, high density fibers tend to be localized in one hemisphere, while communities involving long, low density fibers span the hemispheres.
\item \textbf{Problem 4: Identifying Physical Correlates of Multiresolution Structure} To investigate structure that spans geometric scales, we vary the resolution parameter in modularity maximization to probe mesoscale structure at large (few communities) and small scales (many communities). By calculating community radius, we find that large communities, as measured by the number of nodes, are embedded in large physical spaces.
\item \textbf{Problem 5: Demonstrating Clinical Relevance} To test diagnostic applicability, we apply our methods to functional (fMRI) networks extracted from a people with schizophrenia and healthy controls. Using windowed thresholding, we observe previously hidden significant differences between the two groups in specific weight ranges, suggesting that multi-resolution methods provide a powerful approach to differentiating clinical groups.
\end{itemize}
In the remainder of this Results section, we describe each of these problems and subsequent observations in greater detail.

\subsection{Probing Drivers of Weighted Modularity}
\label{sec:results.power_rew}

We begin our analysis with measurements that illustrate the sensitivity of mesoscale network diagnostics to the edge weight organization and distribution. We employ the weighted modularity $Q_{w}$ to characterize the community structure of the network and ask whether the value of this diagnostic is primarily driven by the organization of strong edges or by the organization of weak edges.

\begin{figure}[!b]
\centering
	\includegraphics[width = .5\textwidth]{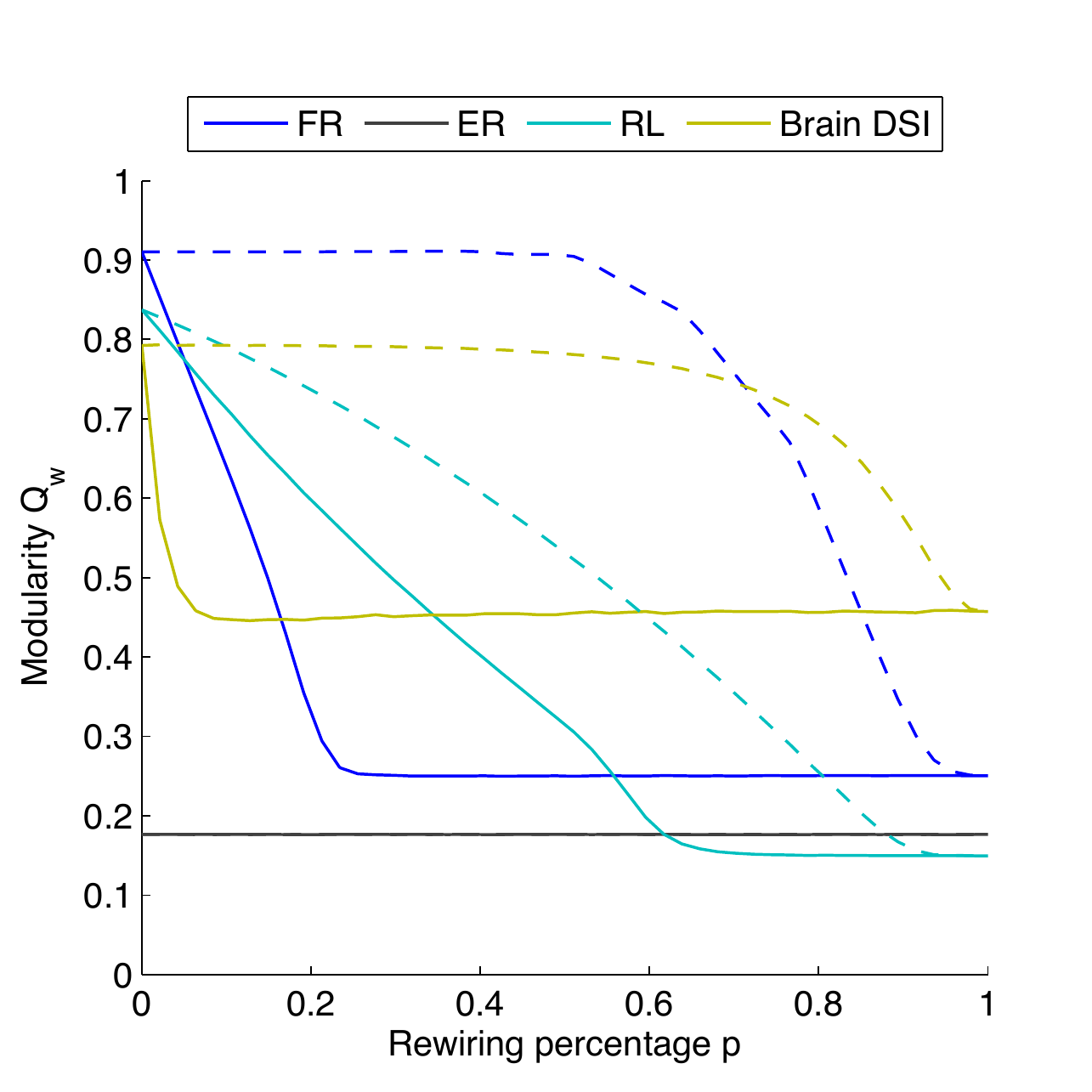}
	\caption{\textbf{Use of Rewiring Strategies to Probe Geometric Drivers of Weighted Modularity} Changes in the maximum modularity as a given percentage $p$ of the connections are randomly rewired for synthetic (fractal hierarchical, \emph{FR}, in blue; Erd\"os-R\'enyi, \emph{ER}, in black; ring lattice, \emph{RL}, in cyan) and empirical (Brain DSI in mustard) networks. Dashed lines show the change of modularity when the $p$ weakest connections are randomly rewired; solid lines illustrate the corresponding results when the $p$ strongest connections are rewired. For the brain data and the synthetic networks, the value of the weighted modularity $Q_w$ is most sensitive to the strongest connections in both the synthetic and empirical networks.}
\label{fig:rewire}
\end{figure}

\paragraph{Rewiring:} By randomly rewiring connections in a weighted network, one can quantify which connections dominate the value of the diagnostic in the original adjacency matrix. In empirical DSI data and three synthetic networks, we compare results obtained when a percentage $p$ of the connections are rewired, starting with the strongest connections (solid lines) or starting with the weakest connections (dashed lines); see Fig.~\ref{fig:rewire}. We observe that modularity $Q_{w}$ as a function of rewiring percentage $p$ depends on the network topology. Randomly rewiring an Erd\"os-R\'enyi network has no effect on $Q_w$ since it does not alter the underlying geometry. However, randomly rewiring a human DSI network or a synthetic fractal hierarchical or ring lattice network leads to changes in $Q_w$, that moreover depend on the types of edges rewired. Rewiring a large fraction of the weakest edges has a negligible effect on the value of $Q_w$, while rewiring even a few of the strongest edges decreases the value of $Q_{w}$ drastically in all three networks. These results illustrate that the value of $Q_{w}$ is dominated by a few edges with the largest weights.

\begin{figure}[!b]
\centering
	\includegraphics[width = .5\textwidth]{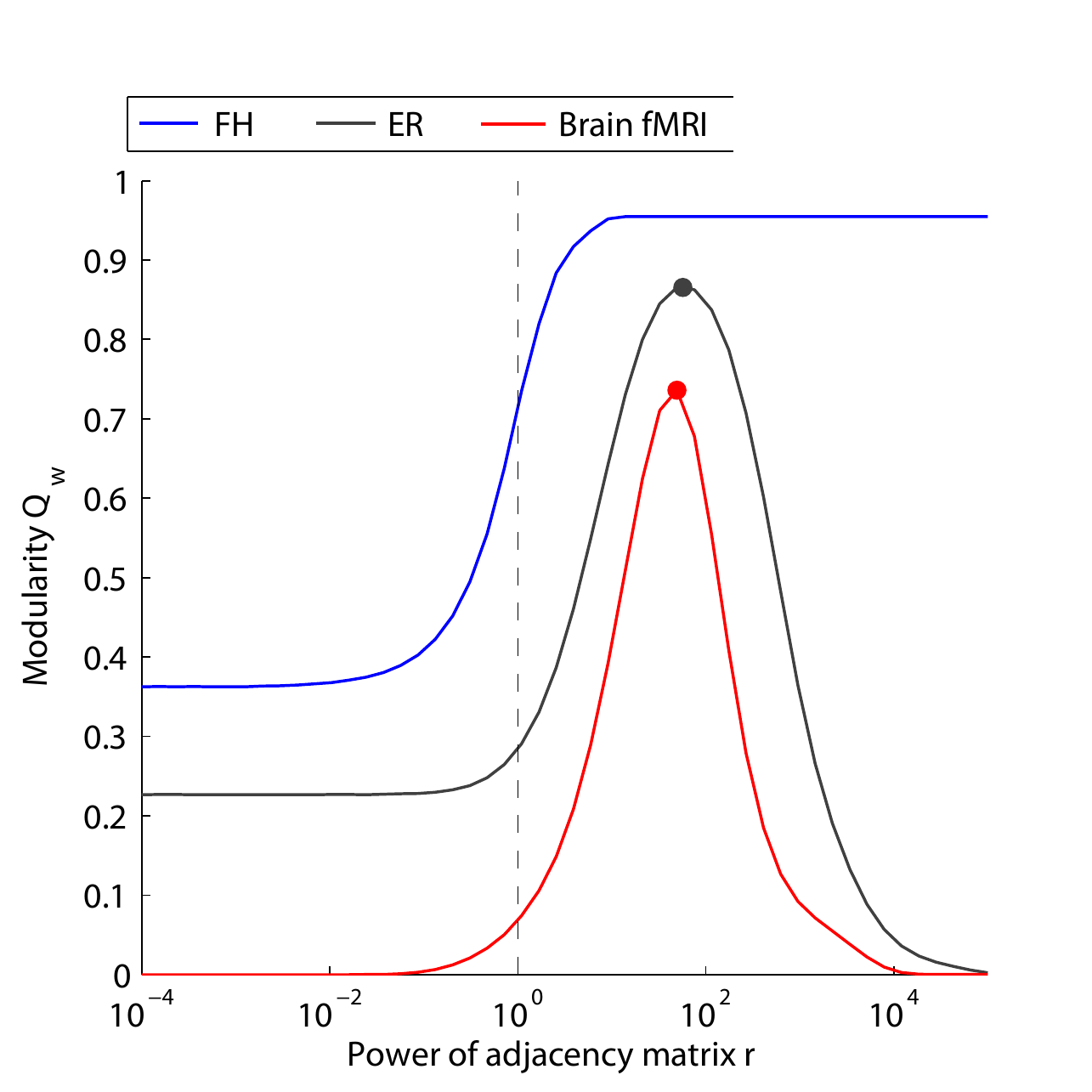}
	\caption{\textbf{Use of Soft Thresholding Strategies to Probe Geometric Drivers of Weighted Modularity} Changes in maximum modularity $Q_{w}(r)$ as a function of the control parameter $r$ for soft thresholding. MRFs are presented for synthetic (fractal hierarchical, \emph{FR}, in blue; Erd\"os-R\'enyi, \emph{ER}, in black) and empirical (Brain fMRI in red) networks. Dots mark the peak value for different curves, which occur at different values of $r$. The vertical dashed line marks the conventional value of $Q_w$ obtained for $r=1$. The single, point summary statistic $Q_w(r=1)$ fails to capture the full structure of the MRF revealed using soft thresholding.}
\label{fig:soft_thresh}
\end{figure}

\paragraph{Soft Thresholding:} To more directly examine the role of edge weight in the value of the maximum modularity, we employ soft thresholding, which individually raises all entries in the adjacency matrix to a power $r$ (see Section \ref{sec:methods.power_and_rew} for methodological details and Fig.~\ref{fig:flowchart} for a schematic). Small values of $r$ tend to equalize the original weights, while larger values of $r$ increasingly emphasize the stronger connections. In Fig.~\ref{fig:soft_thresh} we evaluate the mesoscopic response function (MRF) $Q_{w}$ as a function of $r$, for empirical functional (fMRI) brain networks, as well as two synthetic networks. Our results illustrate that $Q_w(r)$ varies both in shape and limiting behavior for different networks.

In the limit of $r \to 0$, the sparse networks (FH and ER) all converge to a non-zero value of $Q_{w}$, while the fully connected fMRI network converges to $Q_{w}=0$. In the limit of $r \to \infty$, the networks in which all edges have unique weights (i.e., a continuous weight distribution, which occurs in the empirical network as well as the ER network) converge to $Q_{w}=0$, while the FH network, where by construction there exists multiple edges with the maximum edge weight, (i.e., a discrete weight distribution) converge to a non-zero value of $Q_{w}(r)$. Networks with continuous rather than discrete weight distributions display a peak in $Q_w (r)$, but the $r$ value at which this peak occurs ($r_{\mathrm{peak}}$, marked by a dot on each curve in Fig.~\ref{fig:soft_thresh}) differs for each network. The conventional measurement of weighted modularity $Q_w$ is associated with the value for $r=1$ (i.e. the original adjacency matrix), marked by the vertical dashed line in Fig.~\ref{fig:soft_thresh}). In comparing curves it is clear that the conventional measurement (a single point) fails to extract the more complete structure obtained by computing the MRF using soft thresholding.

\paragraph{Summary:} The results presented in Fig.~\ref{fig:rewire} and Fig.~\ref{fig:soft_thresh} provide two types of diagnostic curves that illustrate features of the underlying weight distribution and network geometry. The curves themselves and the value of $r_{\mathrm{peak}}$ in Fig.~\ref{fig:soft_thresh} could be used to uncover differences in multiresolution network structure, for example in healthy versus diseased human brains. However, because the value of $Q_{w}$ is dominated by a few edges with the largest weights, the identification of community structure in weak or medium-strength edges requires an entirely separate mathematical approach that probes the pattern of edges as a function of their weight, such as is provided by windowed thresholding (see next section).

\subsection{Determining Network Differences in Multiresolution Structure}

Determining differences in multiresolution network structure requires a set of techniques that quantify and summarize this structure in mesoscopic response functions of network diagnostics. Windowed thresholding is a unique candidate technique in that it resolves network structure associated with sets of edges with different weights. The technique decomposes a weighted adjacency matrix into a family of graphs. Each graph in this family shares a window size corresponding to the percentage of edges in the original network retained in the graph. A family of graphs is therefore characterized by a control parameter corresponding to the mean weight $g$ of edges within the window. This control parameter can be optimized to uncover differences in multiresolution structure of networks, as we illustrate below.

\begin{figure}[!ht]
\centering
	\includegraphics[width = \textwidth]{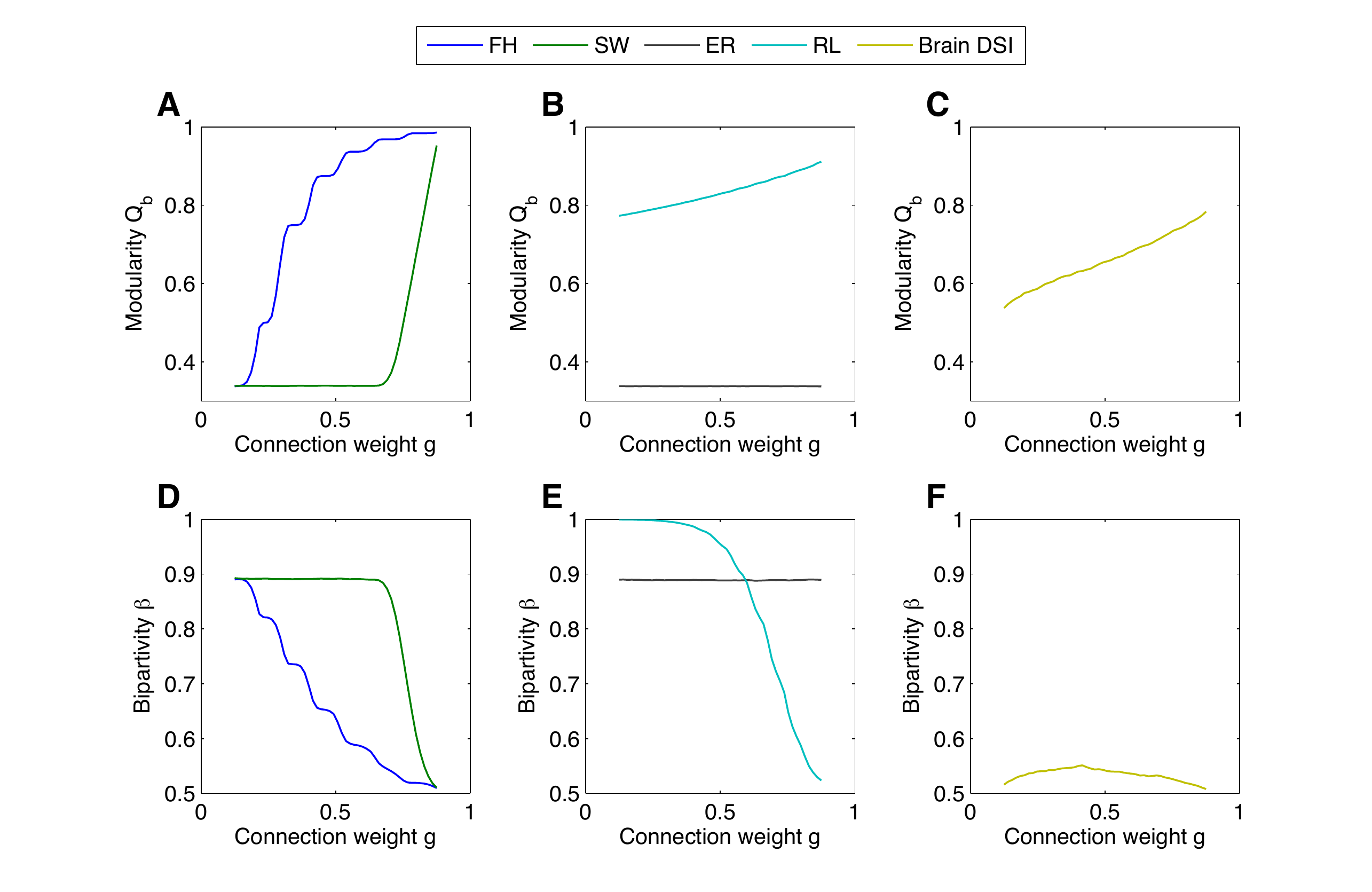}
	\caption{(A-C) Modularity as a function of the average connection weight $g$ (see Sec. \ref{sec:methods.thresholding}) for fractal hierarchical, small world (B), Erd\"os-R\'enyi random, regular lattice (C) and structural brain network (D). The results shown here are averaged over 20 realizations of the community detection algorithm and over 50 realizations of each model (6 subjects for the brain DSI data). The variance in the measurements is smaller than the line width. (D-F) Bipartivity as function of average connection weight of the fractal hierarchical, small world, Er\"os-R\'enyi random model networks and the DSI structural brain networks. We report the initial benchmark results for a window size of 25\% but find that results from other window sizes are qualitatively similar (see the Supplementary Information).
	}
	\label{fig:models.mod.bipart}
\end{figure}

\paragraph{Modularity} The modularity MRF $Q_{b}(g)$ as a function of weight $g$ distinguishes between different networks in both its shape and its limiting behavior (see Figure \ref{fig:models.mod.bipart}A-C). The fractal hierarchical model (FH) yields a stepwise increase in modularity, where each step corresponds to one hierarchical level. The small world model (SW) illustrates two different regimes corresponding to (i) the random structure of the weakest 60\% of the connections and (ii) the perfectly modular structure of the strongly connected elementary groups. The Erd\"os-R\'enyi network (ER) exhibits no weight dependence in modularity, since the underlying topology is constant across different weight values. The ring lattice exhibits an approximately linear increase in modularity, corresponding to the densely interconnected local neighborhoods of the chain. The structural brain network exhibits higher modularity than ER graphs over the entire weight range, with more community structure evident in graphs composed of strong weights.

\paragraph{Bipartivity} The bipartivity MRF $\beta(g)$ as a function of weight $g$ also distinguishes between different networks in both its shape and its limiting behavior (see Fig.~\ref{fig:models.mod.bipart}C-E). The bipartivity of the Erd\"os-R\'enyi network serves as a benchmark for a given choice of the window size, since the underlying uniform structure is perfectly homogeneous, and $\beta (g)$ therefore only depends on the connection density of the graph. The small world model (SW) again shows two different regimes corresponding to (i) the random structure of the weakest 60\% of the connections and (ii) the perfectly modular (or anti-bipartite \cite{Newman2006}) structure of the strongly connected elementary groups. The fractal hierarchical model shows a stepwise decrease in bipartivity, corresponding to the increasing strength of community structure at higher levels of the hierarchy. The regular chain lattice model shows greatest bipartivity for the weakest 50\% of connections and lowest bipartivity for the strongest 50\% of connections. Intuitively, this behavior stems from the fact that the weaker edges link nodes to neighbors at farther distances away, making a delineation into two sparsely intra-connected subgroups more fitting.

\paragraph{Summary} In all model networks, the bipartivity shows the opposite trend as that observed in the modularity (compare top and bottom rows of  Figure~\ref{fig:models.mod.bipart}), supporting the interpretation of bipartivity as a measurement of ``anti-community'' structure \cite{Newman2006}. It is therefore of interest to note that the DSI brain data shows very low bipartivity (close to the minimum possible value of $0.5$) over the entire weight range, consistent with its pronounced community structure over different weight-based resolutions.

\subsubsection{Uncovering Differential Structure in Edge Strength \& Length}
\label{sec:results.brain}

\begin{figure}[!ht]
\centering
\includegraphics[width = \textwidth]{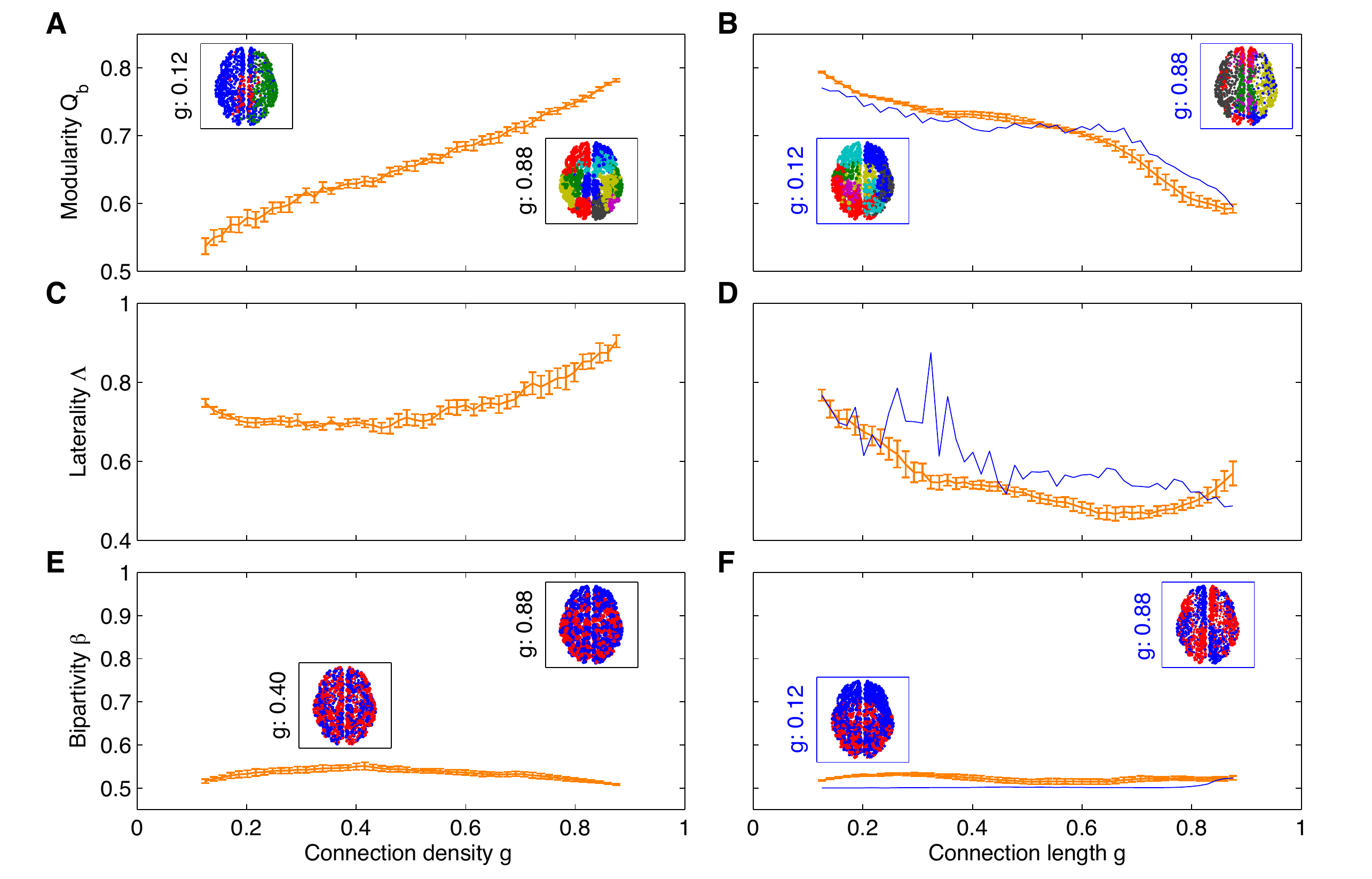}
\caption{ \textbf{Effect of Connection Density and Length on Mesoscale Diagnostics.} \emph{(A-B)} Modularity $Q_b$, \emph{(C-D)} laterality $\Lambda$ and \emph{(E-F)} bipartivity $\beta$ as a function of the fiber tract density density \emph{(A,C,E)}, fiber tract arc length \emph{(B,D,F, blue curve)} and Euclidean distance \emph{(B,D,F, orange curve)}. Orange curves correspond to the mean diagnostic value over DSI networks; blue curves correspond to the diagnostic value estimated from a single individual. The orange curves in panels \emph{B}, \emph{D}, and \emph{F} represent the Euclidean distance between the nodes; the blue curve represents the average arc length of the fiber tracts. The insets illustrate partitions of one representative data set (see Section~\ref{sec:methods}), indicating that communities tend to span the two hemispheres thus leading to low values of bipartivity. All curves and insets were calculated with a window size of $25\%$.}
	\label{fig:brain}
\end{figure}

Unlike the synthetic networks, human brain networks are physically embedded in the 3-dimensional volume of the cranium. Each node in the network is therefore associated with a set of spatial coordinates and each edge in the network is associated with a distance between these coordinates. MRFs for network diagnostics can be used to compare this physical embedding with the network's structure.

To illustrate the utility of this approach, we extract MRFs using two different measures of connection weight $g$ for structural DSI brain data (see Figure~\ref{fig:brain}). In one case $g$ is the density of fiber tracts between two nodes, and in the other case $g$ is the length of the connections, measured by Euclidean distance or tract length. High density and short length edges tend to show more prominent community structure (higher $Q_b(g)$), with communities tending to be isolated in separate hemispheres (higher $\Lambda(g)$). Low density and long edges tend to show less prominent community structure (lower $Q_b(g)$), with communities tending to span the two hemispheres (lower $\Lambda$). Bipartivity peaks for long fiber tract arc lengths, corresponding to the separation of the left and right hemisphere which is even more evident in smaller window sizes (see Supplementary Information). These results suggest a relationship between spatial and topological structure that can be identified across all edge weights and lengths.

The shapes of the MRFs are consistent across individuals. The small amount of inter-individual variability in $Q_b (g)$ and $\beta (g)$ is observed in graphs composed of weak edges. Weak connections may be (i) most sensitive to noise in the experimental measurements, or (ii) most relevant to the biological differences between individuals \cite{Bassett2011a}. Further investigation of individual differences associated with weak connections is needed to resolve this question.

\subsection{Identifying Physical Correlates of Multiresolution Structure}
\label{sec:results.gamma}

\begin{figure}[!ht]
\centering	\includegraphics[width=\textwidth]{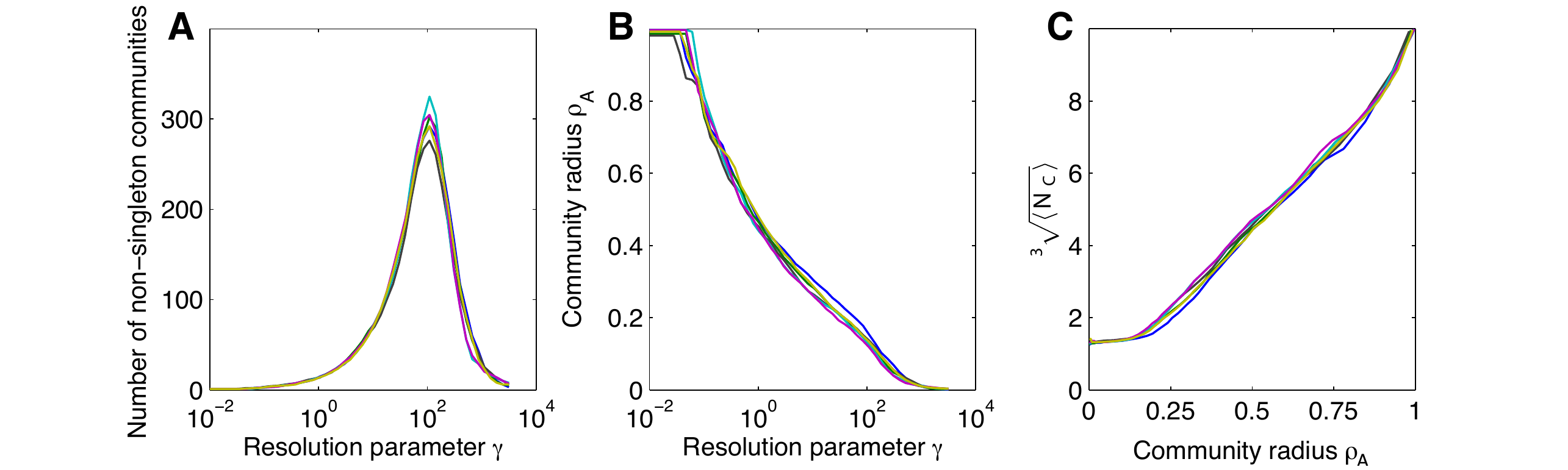}
\caption{\textbf{Effect of the Resolution Parameter on Measured Community Structure.} \emph{(A)} Number of non-singleton communities and \emph{(B)} mean community radius $\rho_A$ as a function of the resolution parameter $\gamma$. \emph{(C)} Mean number of nodes per (non-singleton) community $\mean{N_c}$ as a function of $\rho_A$. Results are presented for networks extracted from the DSI data of six individuals, illustrating consistency across subjects. }
	\label{fig:resolution}
\end{figure}

Windowed thresholding enables us to separately probe the organization of edges according to specific properties that define their weights (e.g., weak versus strong, long versus short, etc.). This prevents strong edge weights from dominating the measurements. However, an important limitation of this method is that network diagnostics are restricted to a particular class of nodes within each window, and thus this method potentially misses important structure associated with the topology connecting different geometrical scales.

To probe community structure at different geometrical scales, we employ a complementary approach. By tuning the resolution parameter $\gamma$ in the optimization of the modularity quality function (Equations 1 and 2), we can identify partitions of the network into both many small (high $\gamma$) and a few large (low $\gamma$) communities (see \ref{sec:methods.gamma}). Compared to our other control parameters, the resolution parameter tunes the output of a diagnostic (e.g. the number of communities), rather than acting directly in the edge weights themselves.

Varying $\gamma$ allows us to probe several features of the network. First, we can uncover the fragmentation profile of a network, as illustrated in Fig.~\ref{fig:resolution}A. For example, in brain DSI networks, the number of non-singleton communities peaks at approximately $\gamma = 100$, after which the network fragments into isolated nodes. Second, we can probe the relationship between community structure and physical network embedding. We observe a polynomial relationship between community radius and the resolution parameter (see Fig.~\ref{fig:resolution}B) and by extension the number of nodes in the community (Fig.~\ref{fig:resolution}C), highlighting the interdependence of geometrical and spatial structure in the brain. Small communities tend to be geographically localized while large communities tend to be geographically distributed, suggestive of efficient embedding \cite{Bassett2010}. Modular networks that are not efficiently embedded into physical space would demonstrate no such relationship. The interaction between space and topology could enhance the organization of information transmission and computing: smaller information processing tasks could be completed by local circuits while larger tasks might make use of more extensive connectivity patterns.

\subsection{Demonstrating Clinical Relevance}
\label{sec:results.groups}

\begin{figure}[!ht]
\centering \includegraphics[width=\textwidth]{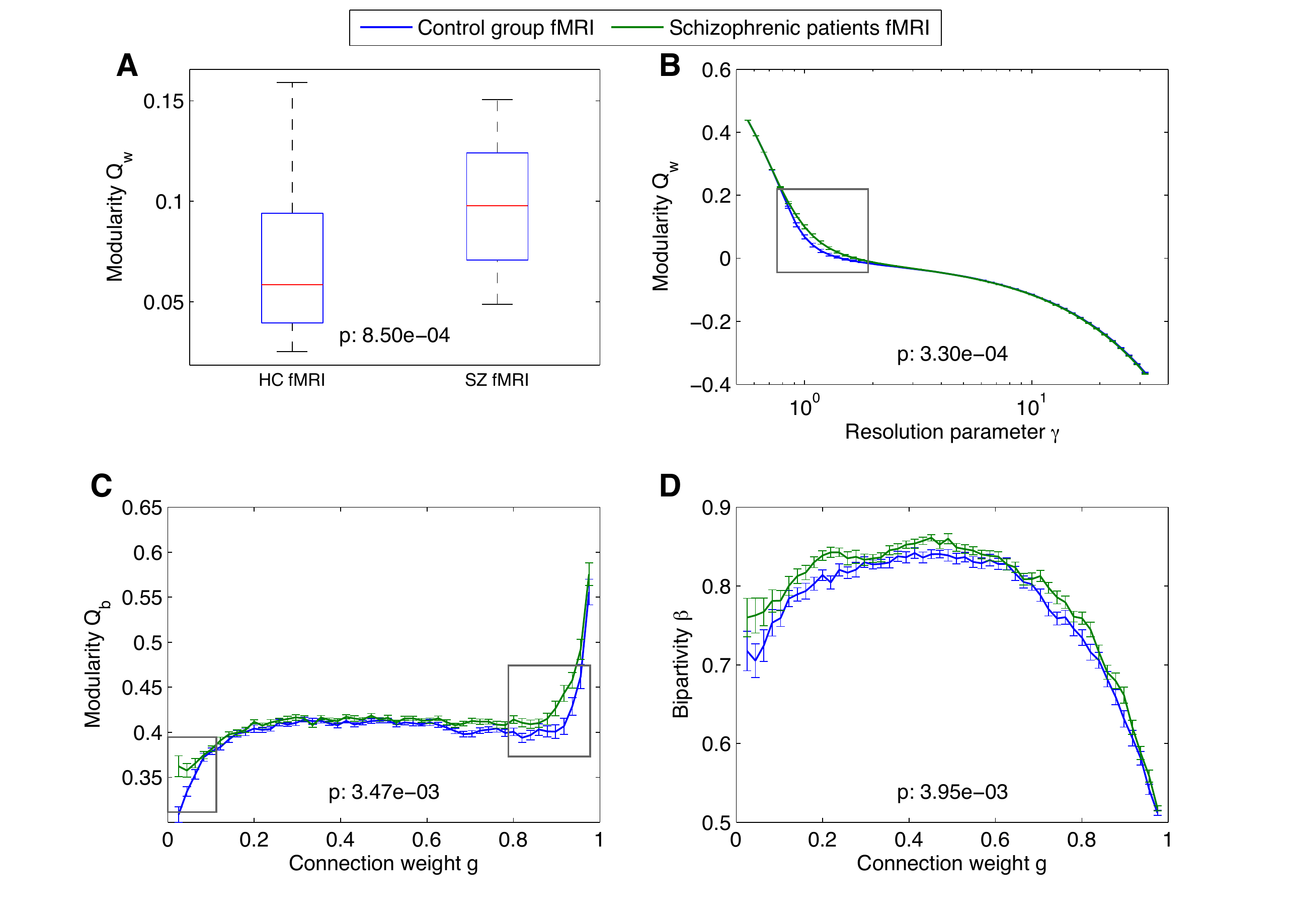}
\caption{\textbf{Multiresolution Mesoscale Structure in Functional Networks.} Functional brain networks were extracted from resting state fMRI data acquired from 29 people with schizophrenia and 29 healthy controls \cite{Bassett2011a} (see Section~\ref{sec:methods}). \emph{(A)} Weighted modularity for healthy controls (left) and people with schizophrenia (right). Box plots indicate range and 25\% (75\%) quartiles over the individuals in each group. The structural resolution parameter is $\gamma = 1$. \emph{(B)} MRFs for the weighted modularity as a function of the resolution parameter $\gamma$. \emph{(C)} MRFs for binary modularity as a function of connection weight. \emph{(D)} MRFs for bipartivity as a function of connection weight. In panels \emph{(C)} and \emph{(D)}, diagnostic values were estimated using a window size of $5\%$. In all panels, $p$-values for group differences in summary statistics (panel \emph{(A)}) and MRFs (panels \emph{(B-D)}) were calculated using a non-parametric permutation test \cite{Bassett2011a}; resolutions displaying the strongest group differences are highlighted by gray boxes. In panels \emph{(B-D)}, error bars show the standard deviation of the mean for healthy controls (blue) and people with schizophrenia (green).}
\label{fig:comp}
\end{figure}

A primary goal of the analysis of human brain networks is to identify changes in network architecture that relate to neurodegenerative diseases or mental disorders. In this section, we investigate the potential applications of multi-resolution techniques to group comparisons by comparing the MRFs for functional brain networks extracted from healthy controls to those extracted from people with schizophrenia.

Functional brain network architecture at rest \cite{Lynall2010,Bassett2011a,Rubinov2009b} and during task performance \cite{Bassett2009,Felix2012,Pachou2008} is known to be altered in schizophrenia \cite{Bassett2009b,Fornito2012a}, supporting the hypothesis that schizophrenia stems from large-scale brain \emph{dysconnectivity} \cite{Stephan2006,Weinberger1992,Volkow1988}. Indeed, we observe that the weighted modularity $Q_w$ (obtained at the default resolution of $\gamma=1$) is significantly higher in people with schizophrenia than it is in healthy controls (nonparametric permutation test: $p\doteq0.001$; see Fig. \ref{fig:comp}A).

MRFs can be used to probe these dysconnectivies in novel ways. By varying $\gamma$, we demonstrate that this group difference is evident over a range of resolutions, corresponding to partitions with both somewhat smaller and somewhat larger communities (nonparametric permutation test of group differences between these curves: $p<0.001$; see Fig.~\ref{fig:comp}B). The MRF also shows that the community structure over very small and very large communities is not different between the two groups, indicating that signatures of dysconnectivity can be constrained to specific resolutions.

To determine whether the patterns of relatively weak or relatively strong edges are most relevant to disease-related alterations in mesoscale brain network architecture, we use the windowed thresholding technique (see Fig. \ref{fig:comp}C). In both groups, graphs composed of strong edges display higher modularity  $Q_b(g)$ than graphs composed of weak edges. However, group differences are predominantly located in graphs composed of either the strongest (high modularity) or weakest (low modularity) edges. No group difference is evident in the bulk of graphs constructed from edges with medium weights, which display modularity values similar to those of the Erd\"os-R\'enyi model at the same connection density. The surprising utility of weak connections in uncovering dysconnectivity signatures has been noted previously in the context of schizophrenia \cite{Bassett2011a} and could potentially be of use in the study of other neuropsychiatric disorders and brain injury.

The organization of weak and strong edges is further elucidated by the MRFs for bipartivity $\beta(g)$ as a function of the connection weight $g$ (see Fig. \ref{fig:comp}D). In both groups, bipartivity values are smallest for edges with large weights. This finding is consistent with those presented in Fig. \ref{fig:comp}C, indicating that strong edges tend to be well localized within functional communities. In general, we observe significantly larger values of bipartivity in people with schizophrenia than in controls (nonparametric permutation test: $p=0.004$). This difference is spread approximately evenly across the entire range of connection weights, in contrast to the group differences in modularity which were localized to a specific resolution range.

\section{Discussion}
\label{sec:discussion}

Functional and structural brain network organization displays complex features such as hierarchical modularity \cite{Bassett2010,Bassett2010c,Meunier2010} and scaling relationships in both topological \cite{Bassett2012f,Bassett2008,Klimm2013} and physical space \cite{Bassett2010}. However, the identification of these structures has largely depended on a binarization of inherently weighted networks. In this paper, we explore several complementary analytical techniques --- including soft thresholding, windowed thresholding, and modularity resolution --- to identify structure at varying scales in weighted brain graphs. We find that brain network structure is characterized by modularity and bipartivity mesoscopic response functions that are shaped unlike those of several synthetic network models. Moreover, the organization of these networks changes appreciably over topological, geometric, and spatial resolutions. Together, our results have important implications for understanding multiresolutions structure in functional and structural connectomes.

\subsection{Multi-Resolution Topological Structure}

Mesoscale structures -- including modularity and bipartivity -- display organization that is dependent on the weight of connections included in the network. The strongest connections in high resolution structural brain networks display stronger community structure, more lateralization of communities to the two hemispheres, and less bipartivity than the set of weakest connections. However, even the weakest connections display a modularity that is greater than expected in a random graph, suggesting the presence of nontrivial structure that might provide important insights into brain organization and function. These results are interesting in light of the fact that such connections have previously been thought to be driven purely by noise and are often removed from the network for statistical reasons \cite{Achard2006}. Our results that weak connections retain structure are consistent with recent evidence demonstrating the potential utility of studying the topology (binary) and geometry (weighted) of weak connections for diagnostic purposes \cite{Bassett2012a}.

Non-random structure in weak connections could stem from multiple factors -- some more biologically interesting than others. First, experimental noise that is preferentially located in particular brain regions (e.g., fronto-temporal susceptibility artifacts) could lead to non-homogeneous network structure. Second, weak connections could be driven by different neurophysiological mechanisms than those driving strong connections (e.g., phase-lagged interactions would be measured as weak connections in correlation-based functional networks). In this case, the mesoscopic response functions would map out a transition between one mechanism for weak connections and another for strong connections. This latter possibility is potentially interesting from a clinical perspective because it could help to disambiguate the role of multiple mechanisms that could drive the altered connectivity patterns evident in many disease states \cite{Bassett2009,Mengotti2011,Sugranyes2012}. In this work, we have simply noted the presence of non-random structure in weak connections but we cannot disambiguate the role of these factors. Further work is necessary to understand the mechanisms driving non-random structure of weak connections.

\subsection{Multi-resolution Spatial Structure}
In addition to weight dependence, mesoscale structures also display organization that is dependent on the length of connections included in the network. Networks composed of relatively long connections show weaker community structure than networks composed of relatively short connections. This distributed nature of long connections is consistent with their hypothetical role in connecting disparate functional modules \cite{Sporns2010}. Furthermore, our results show that communities composed of long connections are more likely to span both hemispheres and display high bipartivity while those composed of shorter connections tend to be more lateralized and display weaker bipartivity, suggesting their role in both inter-hemispheric and inter-lobe communication.

To complement these analyses, we investigated the relationship between community structure and connection distance by employing the structural resolution parameter in the optimization of the modularity quality function. Our results demonstrate that modules composed of few nodes (i.e., community structure at fine-scale structural resolutions) have small spatial radii while those composed of more nodes (i.e., larger-scale structural resolutions) have larger spatial radii. Importantly, this mapping between structural and spatial resolutions would not be expected from a network randomly embedded into physical space \cite{Bassett2010}. Furthermore this relationship between structural resolution and spatial dimension of modules suggests a non-random but rather hierarchical organization within modules, since we find the sub-modules of each module to be have a smaller radius than the super-module. This would not be expected if the sub-structure of a module was randomly organized.

These results highlight the relationship between the geometry of a network (based on edge weights) and the physical embedding of that network into 3-dimensional space, a relationship which is of interest in a wide variety of complex systems \cite{Barthelemy2011}. Such a relationship is consistent with a large body of prior work demonstrating that brain networks extracted from a range of species tend to have near-minimal wiring \cite{Bassett2010,Bassett2010c,Kaiser2011,Kaiser2006,Chen2006,Raj2011}, implicating a density of connections in local geographic neighborhoods and a sparsity of connections bridging those neighborhoods. In contrast to these previous studies that link physical distance to global (path-length and efficiency) and local (clustering coefficient and local efficiency) network diagnostics, our work uncovers the complementary influence of physical space on meso-scale structures (modularity and bipartivity).

The interaction between space and topology in brain systems could be driven by energetic and metabolic constraints on network development \cite{Bullmore2009,Bullmore2012,Bassett2010,Bassett2010c}. Such constraints might also play a role in the fine-grained spatial geometry of white matter fiber tracts, which cross one another at $90$ degree angles \cite{Wedeen2012}, thereby potentially minimizing electromagnetic interference. Moreover, such constraints likely have important implications for system function, where short connections are potentially easier to maintain and use than long connections \cite{Vertes2012}. If such a functional consequence of physical constraints existed, it might partially explain the functional deficits observed in disease states associated with large-scale disconnectivity \cite{Vertes2012,Bassett2009b,Bassett2008,AlexanderBloch2010}. However, converse evidence from normal human development indicates that some distributed processing based on long distance connections is necessary for healthy cognitive function \cite{Fair2009}. Future work is necessary to better understand the role of physical constraints on brain development and organization.

\subsection{Brain Symmetries}
While community structure in functional and structural brain networks has been examined in a number of studies \cite{Meunier2009,Meunier2010,Bassett2010,Chen2008}, other types of mesoscale properties have been less studied. Here we employ two diagnostics -- bipartivity and community laterality -- that could capture signatures of one of the unusual properties of the brain compared to other complex systems: the physical symmetry between the two hemispheres. Our data suggest that such symmetry is observable in the organization of brain networks. The lateralization of communities is greatest for strong, local connections and smallest for weak, long-range connections, consistent with the preference for small modules to contain nodes in a local geographic neighborhood. Conversely, the bipartivity is greatest for mid-strength and long-range connections respectively and the two parts of the bipartite structure appear to map out both anterior-posterior and left-right axes of brain development. The putative functional role of these network symmetries is at least preliminarily supported by our finding that bipartivity of resting state functional brain networks in people with schizophrenia is significantly higher than that in healthy controls. Such whole-brain network signatures could derive from local asymmetries in white matter microstructure \cite{Miyata2012} and decreased interhemispheric white matter connectivity previously observed in schizophrenia \cite{Knochel2012}.

\subsection{Combining Multiresolution Techniques}
In this paper, we have described several complementary techniques for uncovering multi-resolution structure in weighted networks. Each method has advantages and disadvantages and enables one to capture different features present in the network data. Here, for example, we have illustrated these techniques in probing the geometry (weak versus strong edges), embedding (short versus long edges), and structural resolution (small versus large communities) of network architecture. However, in some cases one might wish to probe multiple features of the network architecture simultaneously.

\begin{figure}
\centering
\includegraphics[width=\textwidth]{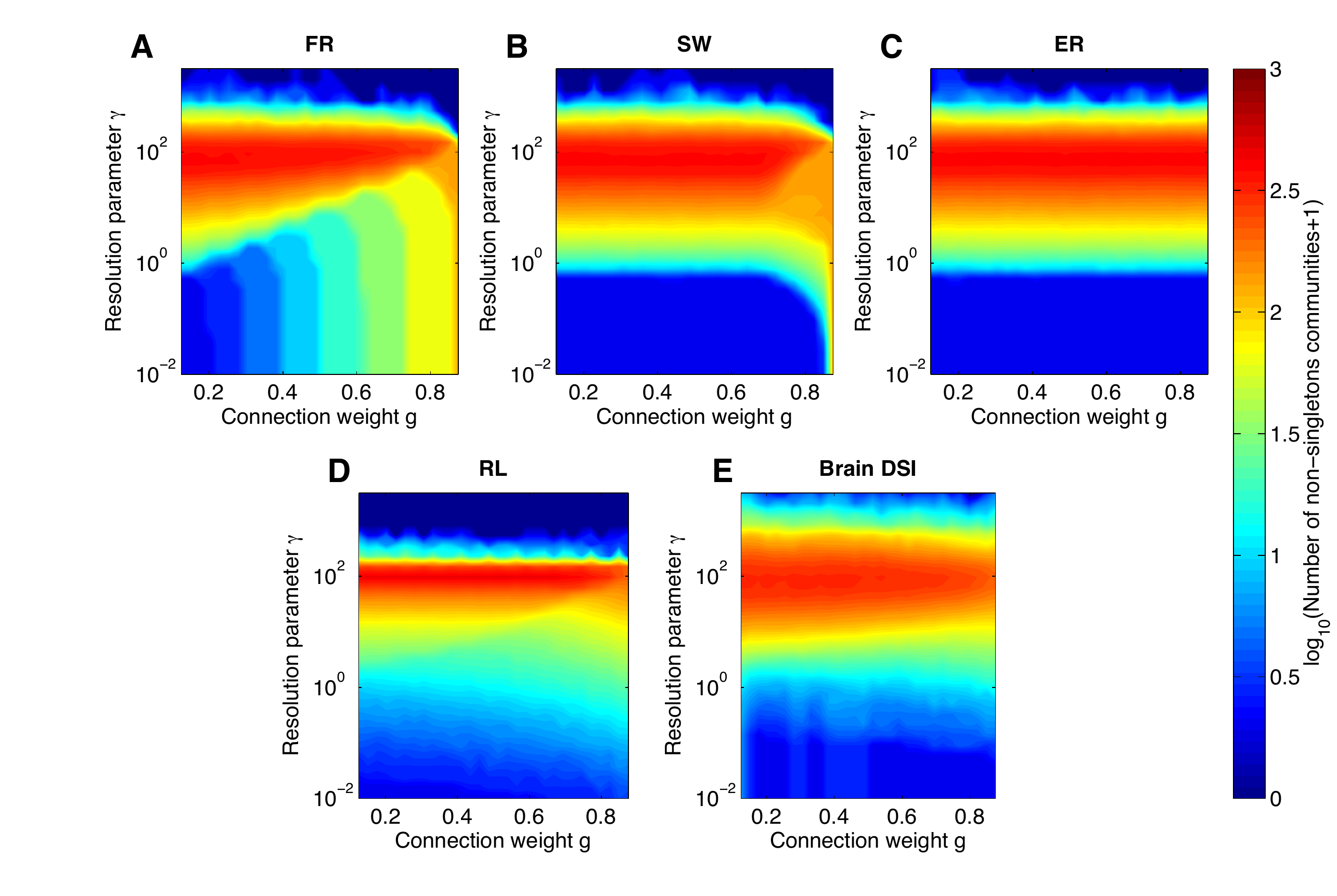}
\caption{\textbf{Simultaneously Probing Structural Resolution and Network Geometry.} Color plots of the number of non-singleton communities as function of both average connection weight $g$ and resolution parameter $\gamma$ for the \emph{(A)} fractal hierarchical, \emph{(B)} small world, \emph{(C)} Erd\"os-R\'enyi, and \emph{(D)} ring lattice, and for \emph{(E)} one representative DSI anatomical network. The window size is 25\%. For results of the total number of communities (singletons and non-singletons), see the Supplementary Information.}
	\label{fig:mod3d}
\end{figure}

In Figure~\ref{fig:mod3d}, we illustrate the combined use of the structural resolution parameter and windowed thresholding to elucidate the fragmentation profiles of empirical and synthetic network models. In the fractal hierarchical network, we observe that the banding of the number of communities as a function of the mean connection weight $g$ is evident for small but not large values of the structural resolution parameter $\gamma$, indicating that the structural organization of the network can be hidden if one studies communities of small size. The remainder of the models and the empirical network data display a relatively constant number of communities over a broad range of mean connection weights for a given value of the structural resolution parameter $\gamma$. This consistency suggests, not surprisingly, that the number of communities is driven more by $\gamma$ than by the underlying network topology, which for each of these networks (with the exception of the Erd\"os-R\'enyi mode) varies over different mean connection weights. Networks completely fragment (zero non-singleton communities at high values of the structural resolution parameter) at similar but not identical values of $\gamma$, making comparisons across topologies for fixed $\gamma$ difficult. We plan to investigate the combined use of methods in more detail in future work.

\subsection{Methodological Considerations}
\paragraph{Models of Brain Structure}
In this work we compared observed multi-resolution organization in brain networks to the organization expected in 4 synthetic networks. While many synthetic network models exist for such a comparison, we chose two benchmark networks with local (ring lattice) and global (Erd\"os-R\'enyi) properties and two networks with mesoscale structures including community structure and hierarchical community structure. Our results show that none of these models displays similar modularity or bipartivity MRFs to those of the brain. An important area of investigation for future work is the generation of alternative synthetic network models that conserve additional network properties of brain systems or employ more biologically realistic growth mechanisms (see for example \cite{Klimm2013}). Moreover, we assigned the edge weights in these models in the simplest possible manner and these models therefore produce weight-dependent changes in topology by construction. Alternative weighting schemes could provide more sophisticated generative models of network geometries that more closely mimic brain structure.

\paragraph{Distance Bias in Anatomical Networks}
The diffusion spectrum imaging network contains an inherent distance bias \cite{Hagmann2008}, meaning that long distance connections have a lower probability of being included in the network than short distance connections. While Hagmann and colleagues did use a distance bias correction in the preprocessing of these networks, the most complete correction method remains a matter of ongoing debate \cite{Hagmann2008}. It is possible that some distance bias remains in the current DSI data sets that might artifactually inflate the observed relationship between space and topology. Indeed, an important area of future work remains to understand the effect of local wiring probabilities (both real and artifactual) on observed network organization \cite{Henderson2011}.

\section{Conclusions}

Our work demonstrates several benefits to the employment of multiresolution network analysis techniques. Such techniques enable statistically meaningful assessments of the organization of weighted spatial networks as a function of edge density, length, and location in Euclidean space. The ability to resolve mesoscale properties -- like modularity and bipartivity -- over spatial and geometric scales facilitates a deeper understanding of network organization than is possible in the examination of any single resolution alone. Moreover, it enables more focused mechanistic hypotheses for altered connectivity profiles in clinical states where network organization might be perturbed in one resolution (weak or local connections) more than in another (strong or long connections).

\clearpage
\newpage
\newpage
\section*{Supplementary Information}

In this Supplementary Document, we including the following materials to support the work described in the main manuscript:
\begin{itemize}
    \item Supplementary Results on Optimization and Realization Variance
    \item Supplementary Results on the Effects of Window Size
    \item Supplementary Results on the Relationship Between the Number of Communities and Singletons.
    \item Table S1: Description of Network Ensembles.
    \item Figure S1: Optimization and Realization Variance.
    \item Figure S2: Effect of Multiresolution Network Geometry on Community Structure.
    \item Figure S3: Effect of Multiresolution Network Geometry on Bipartite Structure.
    \item Figure S4: Role of Singletons in Community Number and Modularity.
    \item Figure S5: Simultaneously Probing Structural Resolution and Network Geometry.
\end{itemize}

\newpage

\subsection*{Ensemble Variability and Optimization Error}
\label{sec:errors}

When comparing two different data sets or models, the interpretation of results can be affected by the relative roles of different sources of variation \cite{Bassett2012}. In this section, we describe the roles of optimization and realization variance.

\paragraph{Optimization Variance} Identifying the optimal partition of a network into communities via modularity maximization is NP-hard. We employ a Louvain-like locally greedy heuristic algorithm \cite{genlouvain} and perform the optimization multiple times to obtain sets of partitions the capture the representative structure in the network. We define the \emph{optimization variance} to be the standard deviation in network diagnostic values over from these multiple optimizations.

\paragraph{Realization Variance} Synthetic network models can be used to produce ensembles of binary or weights graphs based on a given construction or growth rule \cite{Klimm2013} (see Table \ref{tbl:ensembles}). Similarly, empirical networks from neuroimaging data sets represent ensembles of graphs over different human subjects. We define the \emph{realization variance} to be the standard deviation of network diagnostic values over such multiple empirical or synthetic realizations.

\begin{figure}[!ht]
\centering
\includegraphics[width=\textwidth]{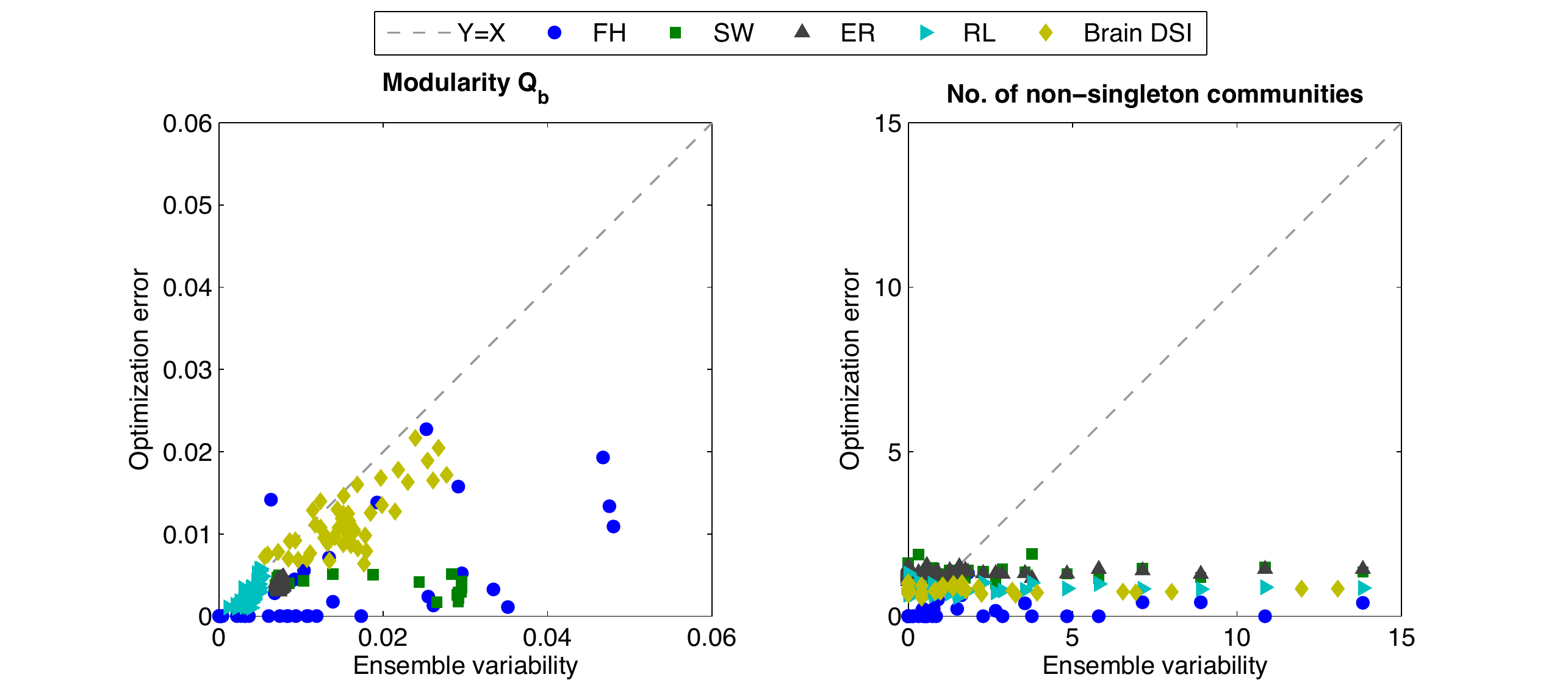}
\caption{\textbf{Optimization and Realization Variance.} The optimization variance versus the realization variance in the \emph{(A)} binary modularity and \emph{(B)} number of communities for the ensembles of fractal hierarchical (blue), modular small-world (green), Erd\"os-R\'enyi (gray), ring lattice (cyan), and DSI brain (gold) networks. The dashed gray line indicates the the line of equivalence between the optimization and randomization variance.}
\label{fig:errors}
\end{figure}

To identify the sensitivity of our results to optimizations and realizations, we determine whether one source of variance is significantly larger in magnitude than another. In general, we find that both types of variance are small in comparison to the quantities of interest such as the modularity or number of communities (see Figure~\ref{fig:errors}). Furthermore, the optimization variance is generally smaller than the realization variance, demonstrating that modularity-based diagnostics are reliable measures of individual variations in mesoscale structures and can be utilized for group comparisons.

\begin{table}[ht]
\centering
\begin{tabular}{lccrcl}
\toprule
Network Type I & Size of ensemble & Nodes & \multicolumn{3}{c}{Edges K $[10^3]$}\\
\midrule
Brain DSI & 6 & 1000 & $14.4$ & $\pm$ & $0.5$ \\
Erd\"os-R\'enyi & 50 & 1000 & $14.4$ & $\pm$ & $0.45$ \\
Regular Lattice & 50 & 1000 & $14.3$ & $\pm$ & $0.6$ \\
Fractal Hierarchical & 50 & 1000 & $14.4$ & $\pm$ & $0.6$ \\
Small World & 50 & 1000 & $14.4$ & $\pm$ & $0.46$ \\
\bottomrule
\toprule
Network Type II\\
\midrule
Brain fMRI & 29 & 90 & $4.05$ & $\pm$ & $0$ \\
Erd\"os-R\'enyi & 50 & 90 & $0.524$ & $\pm$ & $0.025$ \\
Regular Lattice & 50 & 90 & $0.526$ & $\pm$ & $0.026$ \\
Fractal Hierarchical & 50 & 90 & $0.526$ & $\pm$ & $0.030$ \\
Small World & 50 & 90 & $0.523$ & $\pm$ & $0.032$ \\
\bottomrule
\end{tabular}
\caption{\textbf{Description of Network Ensembles.} The number of networks (size) in the ensemble, number of nodes, number of edges $K$ given in units of $[10^3]$ edges for the two types of ensembles studied. Network Type I contains the brain DSI, Erd\"os-R\'enyi, Regular Lattice, Fractal Hierarchical, and Small World networks with $N=1000$ nodes. Network Type II contains the brain fMRI, Erd\"os-R\'enyi, Regular Lattice, Fractal Hierarchical, and Small World networks with $N=90$ nodes.}
\label{tbl:ensembles}
\end{table}

\subsection*{Effects of Window Size}
\label{sec:results.windowsize}
When using the windowed thresholding technique, it is important to understand how the choice of window size affects measured network diagnostic values. Our results are generally robust over a fairly wide range of window sizes between $10\%$ and $25\%$, showing that this method allows a robust analysis of the graph structure. In this section, we provide results obtained with a window size of $15\%$ to complement the results in the main manuscript obtained with a window size of $25\%$.

\begin{figure}
\centering
\includegraphics[width=\textwidth]{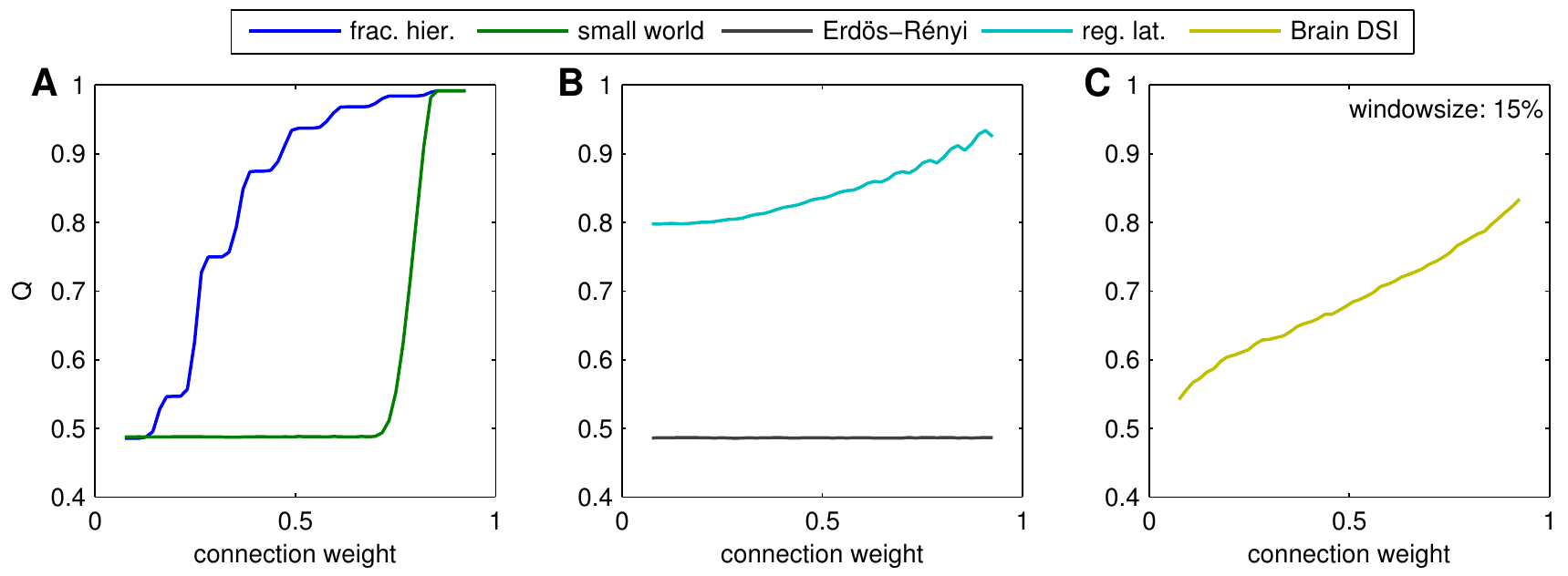}
\caption{\textbf{Effect of Multiresolution Network Geometry on Community Structure.} \emph{(A)} Weighted adjacency matrices depicted for 10\% of nodes in the synthetic network models and structural brain networks extracted from DSI data. \emph{(B-D)} Modularity $Q_b$ as a function of the average connection weight $g$ of the edges retained in the graph (see Section~\ref{sec:methods.thresholding}) for the \emph{(B)} fractal hierarchical, small world, \emph{(C)}, Erd\"os-R\'enyi, regular lattice, and \emph{(D)} structural brain network. Window size is 15\%. Values of $Q_{b}$ are averaged over 20 optimizations of Equation 1 for each of 50 realizations of a synthetic network model or 6 subjects for the brain DSI network. The standard error of the mean is smaller than the line width.}
\label{fig:modularity_ws15}
\end{figure}

Using a window size of 15\%, the MRFs of modularity as a function of connection weight are qualitatively similar to those obtained using a window size of 25\% (compare Figure~\ref{fig:modularity_ws15} to Figure 2 in the main manuscript). The values of modularity obtained using the smaller window size are in general larger than the values obtained using the larger window size, likely due at least in part to the smaller connection density within each graph of the family. Different window sizes can be used to probe different structures in the underlying network geometry. In the small-world model, we are now able to resolve a second plateau in modularity, corresponding to the elementary groups in the construction process. In the fractal hierarchical network, we are also better able to resolve the lowest hierarchical level. In the regular lattice network, we observe small oscillations corresponding to the different lengths of the connections in the chain lattice.

\begin{figure}
\centering
\includegraphics[width=\textwidth]{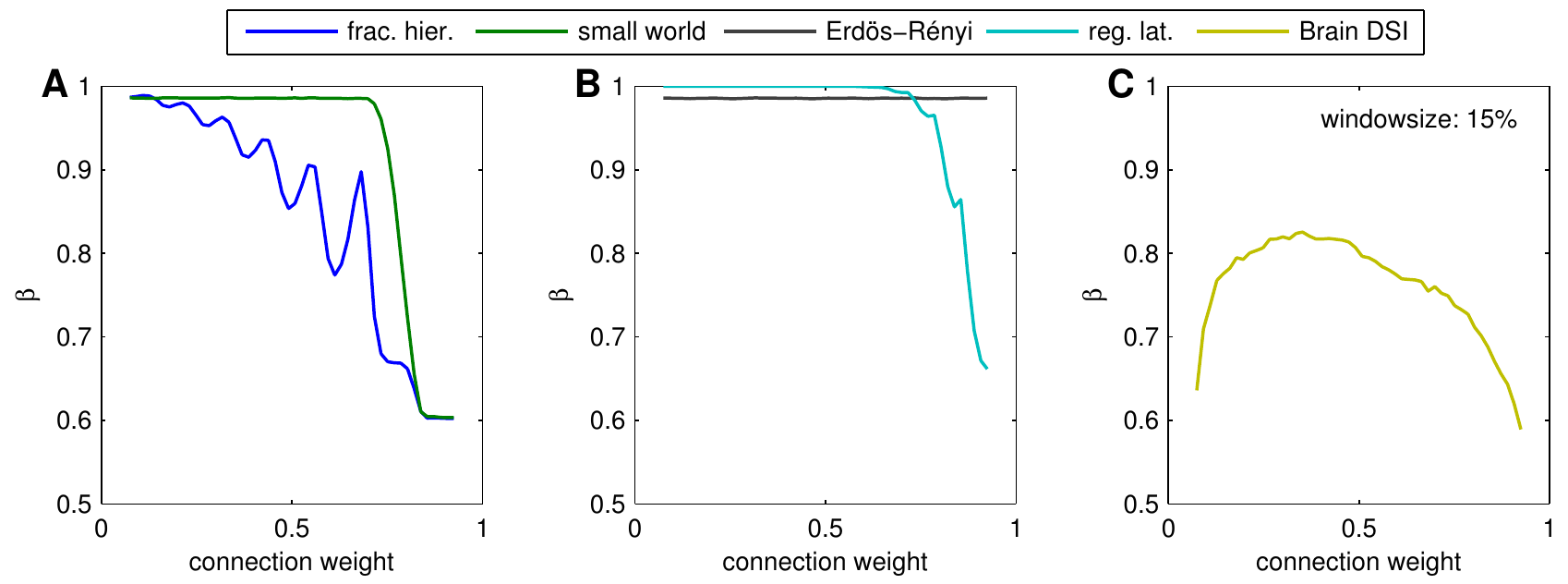}
\caption{\textbf{Effect of Multiresolution Network Geometry on Bipartite Structure.} \emph{(A)} Bipartivity as function of average connection weight $g$ of the edges retained in the graph (see Section~\ref{sec:methods.thresholding}) for the \emph{(B)} fractal hierarchical, small world, \emph{(C)}, Erd\"os-R\'enyi, regular lattice, and \emph{(D)} structural brain network. Window size is 15\%. Values of $\beta$ are averaged over 50 realizations of a synthetic network model or 6 subjects for the brain DSI network. The standard error of the mean is smaller than the line width.}
\label{fig:bipartivity_ws15}
\end{figure}

The MRFs of bipartivity as a function of connection weight obtained using a window of size 15\% are also qualitatively similar to those obtained using a window size of 25\% (compare Figure~\ref{fig:bipartivity_ws15} to Figure 3 in the main manuscript). The values of bipartivity obtained using the smaller window size are in general larger than the values obtained using the larger window size. Similar to the results observed for modularity MRFs, different window sizes can differentially probe networks with different underlying geometries. In the fractal hierarchical model, we observe oscillations that correspond to the transitions between the hierarchical levels. Low bipartivity values correspond to threshold windows containing both inter- and intra-modular edges; high bipartivity values correspond to larger relative numbers of intermodular edges. In the small world model, we are now able to resolve two plateaus corresponding to (i) the elementary groups and (ii) the random organization of the long range connections. For the brain network, we observe that $\beta$ now takes values over a much broader interval.

\subsection*{Number of Communities and Singletons}
\label{sec:results.singletons}

\begin{figure}
\centering
\includegraphics[width=\textwidth]{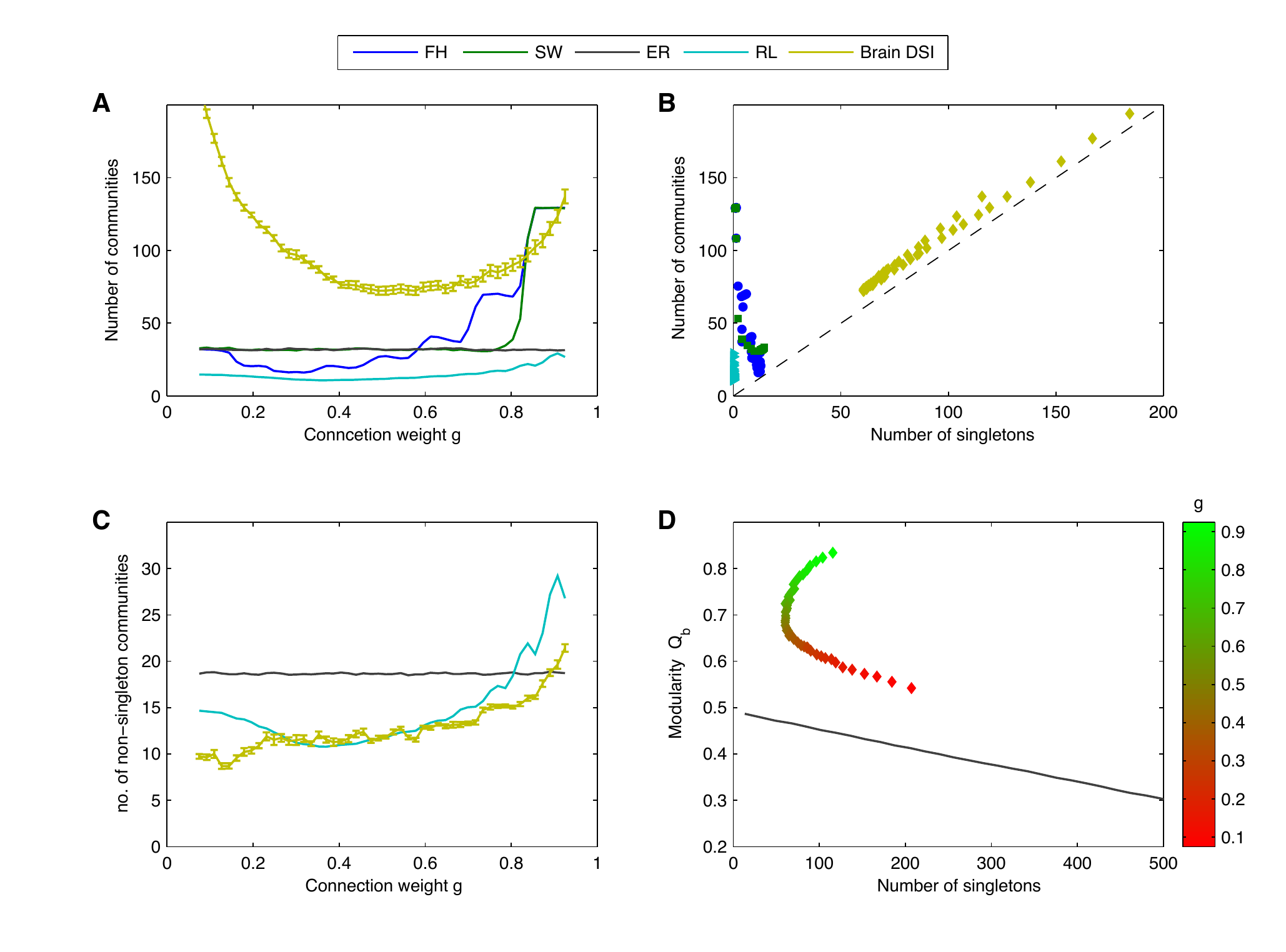}
\caption{\textbf{Role of Singletons in Community Number and Modularity.} \emph{(A)} Number of communities as a function of average connection weight $g$. \emph{(B)} Number of singletons versus number of communities. Data points correspond to each graph in the family which captures network organization at different mean connection weights. \emph{(C)} Number of non-singleton communities as a function of average connection weight $g$. \emph{(D)} Binary modularity as a function of the number of singletons. The gray line shows the modularity of an Erd\"os-R\'enyi random network, when successively disconnecting nodes from the network and randomly adding the same number of connections in the rest of the network. Data points correspond to each graph in the brain DSI family which captures network organization at differen mean connection weights. Color indicates mean connection weight $g$. Window size is 15\%. In panels \emph{(A)} and \emph{(C)}, error bars indicate the standard deviation of the mean; for the model networks this error is smaller than the line width.}
\label{fig:singletons_ws15}
\end{figure}

In the main manuscript, we briefly discussed the fact that when probing community structure across different $\gamma$ values, one can obtain partitions of the network into communities of variable sizes. Singletons are communities composed of a single node. In this section, we examine the role of singletons in two related diagnostics: community number and modularity.

The MRFs of the total number of communities (both singletons and non-singletons) as a function of mean connection weight display different characteristic shapes for different network geometries (see Figure~\ref{fig:singletons_ws15}A). In the Erd\"os-R\'enyi model, the number of communities is constant over the whole range of connection weights, providing a benchmark for model comparison. In the fractal hierarchical network, the different hierarchical levels are evident in the stepwise increases in the community number. In the small-world model, we recover the two distinct structural regimes: (i) the weaker edges display a similar community number to that expected in an Erd\"os-R\'enyi network of the same size, and (ii) the strongest edges display the same number of elementary groups as the fractal hierarchical model.

The brain DSI network contains more communities than the synthetic models, especially at either end of the weight spectrum (i.e., strongest and weakest edges). However, unlike the synthetic models, most of the communities in the brain partitions are singletons (see Figure~\ref{fig:singletons_ws15}B). In fact, we observe an approximately linear relationship between the number of communities and the number of singletons in the brain DSI network that we do not observe in the synthetic network models. This result suggests that the brain displays a relatively simple fragmentation process over different mean connection weights.

The number of non-singleton communities is largest for graphs composed of the strongest edges which we know from the main manuscript display pronounced community structure and smallest for graphs composed of the weakest edges which we know display less pronounced community structure (see Figure~\ref{fig:singletons_ws15}C). In fact, the value of the binary modularity can be directly mapped to the number of singletons (see Figure~\ref{fig:singletons_ws15}D). In the human brain, graph composed of weak edges display small values of modularity and a large number of singletons. Graphs composed of medium-weighted edges display middling values of modularity and a small number of singletons. Graphs composed of strong edges display large values of modularity and a large number of singletons.

To determine whether this complex relationship could be expected, we constructed a benchmark Erd\"os-R\'enyi model and tuned the number of singletons while retaining a fixed connection density using the following algorithm. We created a set of graphs with a different number of singletons. To create a singleton in a given graph, we removed all edges that emanated from a single node $i$ chosen uniformly at random. Let the number of edges removed be denoted $k_{i}$. We then add $k_{i}$ edges to the remaining network, distributing them uniformly at random to all nodes except node $i$. To create many singletons, we iteratively applied the process for creating a single singleton. In this manner, we constructed a set of graphs Erd\"os-R\'enyi graph with between $0$ and $N/2$ singletons.  This process ensures that the number of singletons can be carefully titrated in a graph, the remaining connections still display an Erd\"os-R\'enyi topology, and the connection density of the graph is unaltered.  The modularity value of this Erd\"os-R\'enyi model is approximately linearly related to the number of singletons (see Figure~\ref{fig:singletons_ws15}D). Note that since the singletons are all disconnected nodes and therefore have a zero contribution to the modularity $Q$, independent of the particular partition. Our approach here is mathematically equivalent to decreasing the number of nodes $N$ while keeping the number of connections $K$ fixed. In this sense one can say that the modularity scales linearly with $N$ over a fairly wide range of the percentage of singletons. This is consistent with the observation of decreasing modularity for higher numbers of connections across the ensemble of benchmark Erd\"os-R\'enyi networks.

Thus, the complex relationship between modularity and number of singletons observed in the human brain is not expected from an benchmark Erd\"os-R\'enyi graph. This is perhaps unsurprising given that the graphs in the human brain network family each contain inherently different topologies while the benchmark Erd\"os-R\'enyi graphs are constructed to retain the same topology as the number of singletons increases. The results highlight the difficulties in interpreting the actual value of the modularity itself, and instead support efforts in utilizing other diagnostics based on the partition structure \cite{Bassett2012}.

\begin{figure}
\centering
\includegraphics[width=\textwidth]{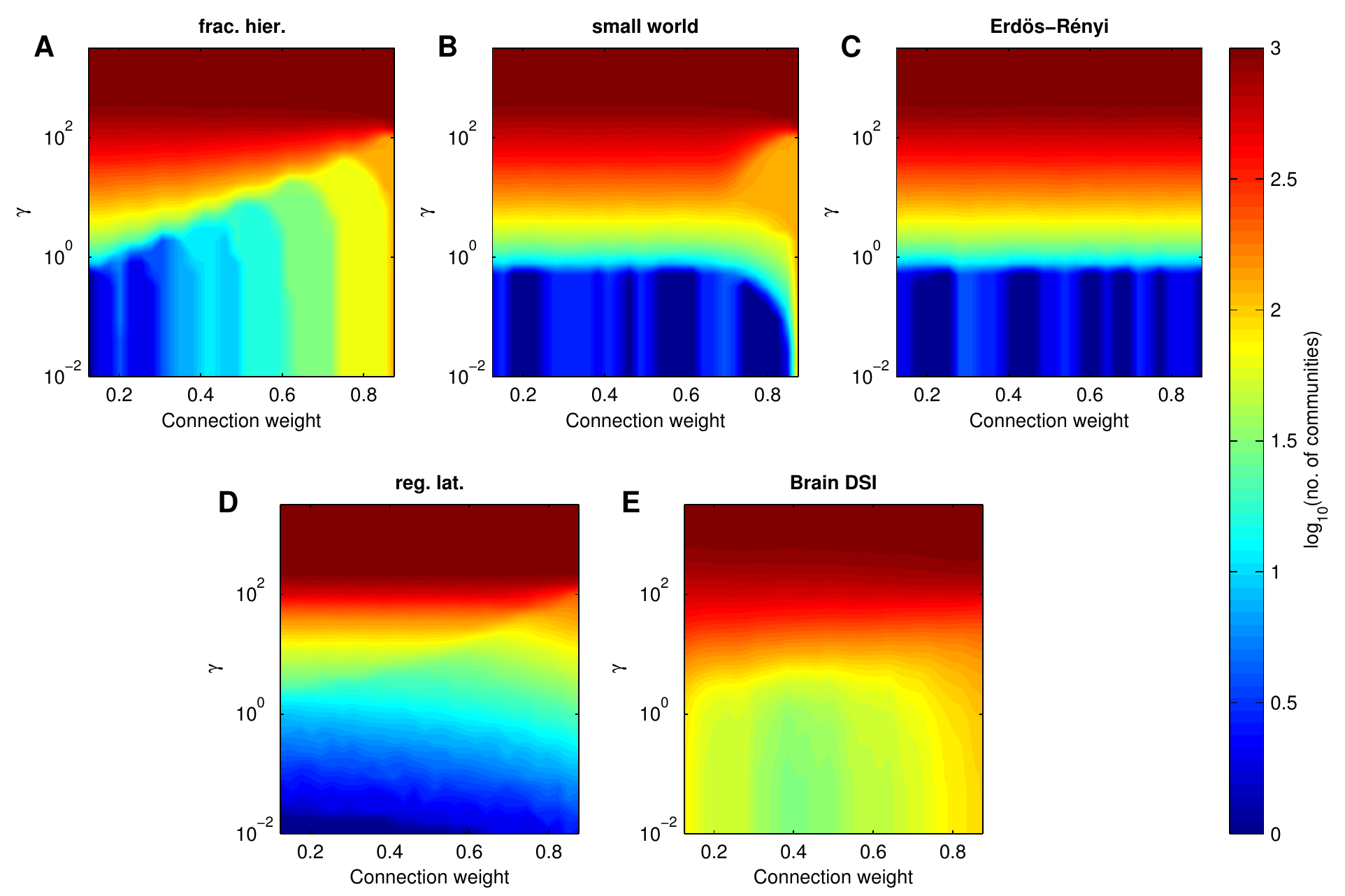}
\caption{\textbf{Simultaneously Probing Structural Resolution and Network Geometry.} Colorplots of the total number of communities (singletons and non-singletons) as function of both average connection weight $g$ and resolution parameter $\gamma$ for the \emph{(A)} fractal hierarchical, \emph{(B)} small world, \emph{(C)} Erd\"os-R\'enyi, and \emph{(D)} regular lattice models and for \emph{(E)} one representative DSI anatomical network. The window size is 25\%.}
\label{fig:comm3d}
\end{figure}

\bibliographystyle{plos2009}
\bibliography{Bibliography_DB}

\end{document}